%% file: cidor.tex
\documentclass[10pt,conference]{IEEEtran}

\usepackage{url}
\newcommand{\rom}[1]{\romannumeral#1\relax}

\usepackage{amssymb}

\usepackage{algorithm}
\usepackage{algorithmic}

\usepackage{graphicx}
\usepackage{subfig}

\usepackage{amsthm}

\theoremstyle{Plain}

\usepackage{array}

\begin{document}




\title{CIDOR: Content Distribution and Retrieval in Disaster Networks for Public Protection}

\author{\IEEEauthorblockN{Hasan M A Islam$^1$, Dmitrij Lagutin$^1$, Andrey Lukyanenko$^1$, Andrei Gurtov$^2$ and Antti Yl{\"a}-J{\"a}{\"a}ski$^1$} 
\IEEEauthorblockA{$^1$Aalto University, $^2$Link{\"o}ping University \\
Email: $^1$firstname.lastname@aalto.fi, $^2$firstname.lastname@liu.se
} 
\vspace{-0.3cm}
}

\maketitle

\begin{abstract}
Information-Centric Networking (ICN) introduces a paradigm shift from a host centric communication model for Future Internet architectures. It supports the retrieval of a particular content regardless of the physical location of the content. Emergency network in a disaster scenario or disruptive network presents a significant challenge to the ICN deployment. In this paper, we present a Content dIstribution and retrieval framework in disaster netwOrks for public pRotection (CIDOR) which exploits the design principle of the native CCN architecture in the native Delay Tolerant Networking (DTN) architecture. We prove the feasibility and investigate the performance of our proposed solution using extensive simulation with different classes of the DTN routing strategies in different mobility scenarios. The simulation result shows that CIDOR can reduce the content retrieval time up to 50\% while the response ratio is close to 100\%.


\end{abstract}










\section{Introduction}

The information centric networking (e.g., CCN \cite{jacobson2009networking}, DONA \cite{koponen2007data}, PURSUIT/PSIRP \cite{tarkoma2009publish}, POINT \cite{point}, NetInf \cite{dannewitz2009netinf}) emerges as a paradigm shift from the host centric conventional Internet. The conventional Internet is becoming increasingly infeasible and less meaningful in terms of a number of recognized limitations, for instance, decoupling address from an end-point identity, mobility and disruption tolerance, and above all, scalable and efficient content distribution. The extensive usage of current Internet is dominated by the content distribution and retrieval of a substantial amount of digital content. ICN allows user to retrieve a particular content regardless of any reference to the physical location of the content. However, the emergency network, such as the consequence of a natural disaster scenario like earthquake, hurricane, or tsunami, or network disruption means a significant challenge for the ICN deployment. For instance, name resolution may fail due to network disruptions, especially when the components of the distributed resolution services are devastated by network partitioning. Moreover, name resolution may become unreliable, particularly when content names are resolved to locator(s) that do not exist anymore. In such scenario, it is very essential to spread the important news for public protection and disaster relief. Nevertheless, in such scenario mobile devices can be exploited to form a peer-to-peer network. This issue is addressed in the Delay-tolerant networking (DTN) architecture \cite{cerf2007delay} which follows the store-carry-and-forward model.

The goal of this paper is introducing a novel approach for constructing an Information Centric Networking (ICN) for a disaster network. For such network, we envision the possibility of exploiting the basic design principle of the native Content Centric Networking (CCN) \cite{jacobson2009networking} in DTN architecture. While designing our proposed architecture, we leverage the inherent properties of the CCN and the Bundle Protocol (BP) \cite{burleigh2007bundle} of the DTN architecture. We position our work in the forwarding plane so that the existing DTN routing or the content based DTN routing ( e.g., \cite{neves2013cedo}, \cite{krifa2012message}) can easily be adapted/operated on top of the CIDOR.

In summary, the main contributions of this paper are as follows:
\begin{itemize}

\item We design a content-centric architecture for a disaster network called CIDOR, which includes content-centric framework on top of the Bundle Protocol (BP) \cite{burleigh2007bundle} of the DTN architecture \cite{ccnx} (Figure~\ref{CIDOR_node}). 

\vspace{.15cm}

\item A key contribution of this work is introducing a new forwarding mechanism for DTN based on a new data structure called as CIDOR-PRIT (Section ~\ref{pit}) which stores the pending requester information on the interest packet. 

\vspace{.15cm}

  \item \textit{Duplicate suppression} and \textit{Redundancy elimination} mechanism at the content level (Section~\ref{duplicate} and~\ref{re}) using the CIDOR-PRIT and a randomly generated \textit{Nonce} value which is a part of the request packet. These duplicates may result from the forwarding loop or similar requests issued by multiple requesters. These duplicates can not be detected by the host-centric DTN. 
  \vspace{.15cm}
  
  \item To enable content centric fashion for the DTN and duplicate suppression, the CIDOR utilizes the extension blocks of the DTN message (Section~\ref{format}). The content centric functionality of the CIDOR enables the host centric DTN to perform in a content-centric fashion.
 
  \vspace{.15cm}
  
  \item We have implemented our design as a proof-of-concept in the ONE \cite{keranen2009one} simulator and evaluated thoroughly the efficiency of the CIDOR architecture using simulation with different classes of the DTN routing strategies in different mobility scenarios. The DTN routing strategies are applied on top of the CIDOR architecture.
\end{itemize}

\vspace{.15cm}

The rest of this paper is organized as follows: We first introduce the background in Section~\ref{back}. Section~\ref{motivation} defines the problem statement and gives an overview of our solution. Section~\ref{design} describes our proposed architecture and the implementation details. After that, the proposed architecture is evaluated in Section~\ref{result}, with different classes of DTN routing strategies for the different mobility scenarios. Section~\ref{related} investigates the related work and finally, Section~\ref{conclude} concludes the paper.

\input{motivation}

\input{design}

\input{performance}

\section{\textbf{Related Work}}
\label{related}

\subsection{Mobile Ad-hoc CCN}

In the last few years, numerous research efforts explored the applicability of the CCN in dynamic environments \cite{meisel2010ad, anastasiades2014opportunistic, meisel2010listen, amadeo2011chanet, varvello2011design, strayer2013cascade}. The authors of \cite{varvello2011design} investigates the potential of existing MANET routing algorithms into the CCN based on analytical models. In \cite{anastasiades2014opportunistic}, the authors implement an extension for the CCN caching so that the user can retrieve the content with resume capability in a situation where the data transmission would not have completed. Similar to the CCN, Mooi et al. \cite{chuah2011secure} describes a secure content-centric mobile network (SECON) which enables a user to issue requests in a mobile environment. The authors of \cite{meisel2010listen} has proposed Listen First, Broadcast Later (LFBL) for MANET in line with the named data that is not relying on a predetermined end-to-end path information, the IP addressing, or a MAC layer. In \cite{amadeo2011chanet}, the authors exploit the CCN communication model on top of IEEE 802.11 protocol in the MANET. Similar aspiration is explored in the CASCADE \cite{strayer2013cascade} for the tactical MANET that utilizes the concept of topological and interest-based communities to serve the content quickly in a resource-friendly fashion. Sourlas \textit{et al.} \cite{sourlasinformation} extend the CCN router by introducing a new data structure called Satisfied Interest Table (SIT) which stores the information of the users to whom the data packets are forwarded. In case the server of origin is not reachable, the proposed scheme exploits the SIT entries to forward the request. However, the proposed scheme performs well only if the users listed in the SIT entries are connected. All these works are still based on the implicit assumption of eventual end-to-end connectivity for a while. In contrast, our proposed scheme exploits the opportunistic communication of mobile users using DTN mechanisms (store-carry-and-forward).

\subsection{Combining ICN and DTN}

There are also several research efforts in the DTN environment \cite{tysontowards}, \cite{trossen2016towards}. In \cite{tysontowards}, the author investigates the possibility of integrating the ICN and the DTN principles into a shared ICDTN architecture. Combining the ICN and the DTN has been demonstrated in a recent effort called RIFE architecture \cite{trossen2016towards}. The RIFE is a universal communication architecture that combine the publish/subscribe based POINT architecture \cite{point} and the DTN that provides services for the existing IP-based protocols (e.g., HTTP, CoAP, basic IP) through the ICN core. The IP endpoints are connected through the ICN using a gateway. In contrast, the CIDOR integrates a new content centric framework in DTN, where an end user issues a request based on the CCN naming scheme. The name based replication system \cite{psaras2014name} based on the message priority in a fragmented network can benefit the CIDOR since the system spread important messages quickly and stay longer in the network.

\subsection{Content centric routing for DTN}

User-centric content distribution in the DTNs has been widely explored from various different points of view~\cite{gao2011user, costa2008socially, yoneki2007socio, lu2014information}. Authors of \cite{gao2011user} exploits the caching of mobile users in sharing content items with their neighbours in the same network domain. From a social-based point of view, authors of SocialCast \cite{costa2008socially} proposed a routing framework that exploits the social ties among users for effective relay selection, while Yoneki et al. in \cite{yoneki2007socio} proposed a publish-subscribe based communication overlay maintaining the social groups based on centrality measures. However, this routing mechanisms can be complementary to our proposed scheme which operates independently of any routing algorithm. Lu \textit{et al.} at \cite{lu2014information} used the K-means clustering algorithm to create the social level forwarding scheme for reducing the transmitted messages. This approach raises several inevitable limitations: \textit{(i)} the interest may fail to reach the encountered node with the same social level, that might have the content to satisfy the interest, \textit{(ii)} the request from the higher social level will never reach a content provider with lower social level, \textit{(iii)} the proposed scheme cannot detect the routing loop of the interest packet and, \textit{(iv)} the authors do not consider how to optimise similar interests from multiple users. These limitations are addressed in our solution.

\section{Conclusion}
\label{conclude}

In this paper, we have proposed and investigated a content distribution and retrieval framework (CIDOR) based on the Bundle Protocol (BP) of the DTN architecture. This new architecture achieves the content retrieval, caching and forwarding of the packets more efficiently and enables BP to operate in a content centric fashion in a disaster scenario. The CIDOR introduces a new duplicate suppression mechanism and redundancy elimination technique at content level for BP. Then, we have simulated and recorded the performance of the CIDOR. Next, we have thoroughly studied its efficiency applying the different kinds of DTN routing strategies in different mobility models. The result of our study validates that CIDOR can benefit BP while operating in a content centric fashion. For instance, CIDOR can achieve the response ratio almost 100$\%$ with reduced latency up to 50$\%$ (Figure~\ref{helsinki1},~\ref{rwp}). We also observe that CIDOR enables the hybrid routing \textit{EP$_{p}$SWRouting} (Section~\ref{dr}) to achieve the similar performance compared to Epidemic routing in terms of response ratio with higher delivery probability and much lower cost. While CIDOR bridges the content-centric and host-centric paradigms, it also remains flexible in adapting the existing content-centric DTN routing algorithms because of its modular design.

Our next steps will be focused on the integration of the CIDOR architecture with the infrastructure network which runs the native CCN as well as POINT architecture. We envision that the integration of CIDOR architecture with the native CCN will enrich the connectivity options of the native CCN in a fragmented network. Moreover, we plan to consider the incentive mechanism that motivate the mobile users to cooperate and store others content.

\section{Acknowledgement}
The work presented in this paper was supported by the EU funded H2020 ICT project POINT, under contract 643990. Andrei Gurtov was supported by the Center for Industrial Information Technology (CENIIT)

\tiny
\bibliographystyle{unsrt}
\bibliography{biblio}

\end{document}

%% file: motivation.tex
\section{\textbf{Background}}
\label{back}

\subsection{Content-Centric Networking}
\label{ccn}
Among all the ICN proposals, the CCN is gaining more and more interest for its architectural design. The CCN supports two types of messages: \textit{Interest} and \textit{Data}. Each CCN node maintains three data structures; \textit{Pending Interest Table (PIT)}, \textit{ Forwarding Information Base (FIB)} and \textit{Content Store (CS)}. The CCN communication is consumer driven, i.e., a consumer issues an \textit{Interest} packet towards the content source based on the information stored in the FIB. Upon reception of an interest, a node first checks its local cache for the matching content. Otherwise, the node forwards the \textit{Interest} packet to the interface(s) based on the FIB table until the \textit{Interest} packet reaches a content source. Intermediate nodes store the interests in the PIT table which is used to send the data back to the appropriate requester. In addition, the PIT is used to detect the duplicate interests and suppress the forwarding the duplicates over the same interface. Furthermore, PIT provides content aggregation on a particular node. CCN interest, which is not satisfied within a reasonable amount of time, is retransmitted. As CCN senders are stateless \cite{jacobson2009networking}, the consumer is responsible for re-expressing interest, but only if the interest is not satisfied.

\subsection{Delay Tolerant Networking}

The Delay-tolerant networking (DTN) \cite{cerf2007delay,fall2003delay} is an initiative that introduces an architecture for a challenged environment which is particularly specified by long delay paths, sporadic connections, and network partitions. To achieve this, the DTN creates an  opportunistic network on top of the existing underlying Layer 2 and Layer 3 protocols. This is achieved through an asynchronous communication along with the use of underlying Convergence Layer Adapter (CLA) (TCP, UDP, Bluetooth, etc.). It weakens the necessity of a stable connection between a source and destination end-points for a communication session. Consequently, DTN architecture provides a flexible and resilient protocol for such networks. The DTN is based on store-and-forward model utilizing persistent storage that is well distributed throughout the network. All data are cached in the network until an opportunistic contact is available to forward data. Nevertheless, content-based routing has been explored in the DTN architecture \cite{costa2006adaptive}. The DTN architecture has some properties (e.g., in-network caching, late binding) similar to ICN design and vice versa.

\section{\textbf{Problem Statement}}
\label{motivation}

Similar to IP, the CCN's network layer is designed to operate on unreliable packet services, i.e., it makes a weak demand on layer 2 (e.g., stateless, unreliable, unordered, best effort delivery). Consequently, \textit{Interest}, \textit{Data}, or both might be lost during transport. Unlike TCP, CCN consumers (the application that originates interest) are stateless and are responsible for reissuing an unsatisfied interest. Intermediate forwarders are responsible for retransmission on a particular interface since the forwarder node knows the lifetime of the CCN packet for the upstream node(s). In a emergency network during natural disaster, mobile users are highly dynamic and the connections are intermittent. It is quite difficult to keep track of the network topology change. In such a scenario, retransmission and re-expression of interests might happen due to a large RTT. The retransmission and re-expression of interests may create a redundant network traffic and consume a significant amount of bandwidth. Furthermore, the PIT table may overflow with frequent disruptions that may lead to the unnecessary retransmissions since the previous hop information stored in the PIT table may not exist due to the sporadic connectivity. Nevertheless, the PIT bottleneck may raise an inevitable constraint in terms of reliability and scalability. Thus, a reverse path based on the PIT and interest aggregation are not suitable for the fragmented network. To handle this, we introduce a separate PIT table for the DTN environment called as CIDOR-PRIT. Unlike CCN, the CIDOR-PRIT keeps track of the requester information on the \textit{Interest} packet instead of the arrival interface. Section~\ref{duplicate} and Figure~\ref{response} illustrate the CIDOR operations on the CIDOR-PRIT table and how the CIDOR reduces the generation of redundant \textit{Interest/Data} packets.

In contrast to CCN, the DTN architecture provides a flexible and resilient protocol through asynchronous communication between two end-points, along with the use of underlying Convergence Layer Adapter (CLA) in a fragmented network. 
The fundamental principle of the DTN architecture provides a sender initiated host-centric unicast communication model and still relies on the conventional addressing scheme of the senders and the receivers. To address this, the CIDOR extends the DTN message format to include CIDOR metadata information and enable BP to operate in a content centric fashion. However, The host-centric DTN has no way to detect request forwarding loop at the content level. In addition, messages in the DTN routing are typically identified by the pair of the source and destination addresses and assigned a unique identifier by the originator of these messages. Therefore, messages from different requesters/responders are considered as different ones and hence DTN routing cannot suppress those messages as duplicates. To overcome this, the CIDOR architecture introduces \textit{Duplicate Suppression} mechanism at the content level to detect the forwarding loop and duplicate messages. Moreover, the CIDOR can reduce redundant packet generation by maintaining the CIDOR-PRIT table while operating in a disruptive scenario (see Section~\ref{duplicate} and Figure~\ref{response} for details).


%% file: design.tex
\section{\textbf{CIDOR Architecture}}
\label{design}

\begin{figure}[!t]
\centering
\includegraphics[width=.9\linewidth]{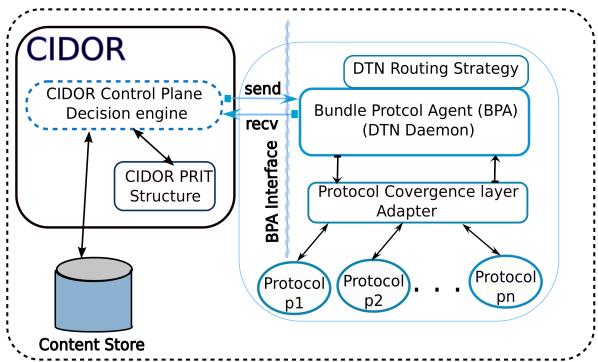} 
\caption{CIDOR Node Engine.}
\label{CIDOR_node}
\end{figure}

CIDOR architecture (Figure~\ref{CIDOR_node}) provides the content-centric framework as an application logic on top of the Bundle Protocol and requests transmission of, accepts delivery of, and processes the CIDOR specific data. The key component of CIDOR architecture is the control plane decision engine that performs packet (Interest/Data) management. The control plane is implemented on top of Bundle Protocol (BP) and its functionalities are responsible for performing specific actions based on the packet type (Interest/Data). For this the control plane inserts the meta-information in DTN messages, that enables the host centric DTN to perform in content centric fashion.

\subsection{CIDOR Routing and Forwarding}
\label{rf}

In the Internet architecture, we differentiate between routing and forwarding. Forwarding is the basic method for transferring the packet to the next hop, i.e., a packet is transferred between a source interface and a destination interface. In contrast, routing is the process by which one router sends packets to another router by means of routing protocols which decide the appropriate path for the packet. The routing protocol assists the router in choosing the best path out of many paths. Nevertheless, the CIDOR is designed in a way that it can operate independently of the DTN routing protocols and therefore complements the existing routing protocols suitable for a particular environment.

\subsection{Packet format}
\label{format}

The CIDOR follows the message format of the Bundle Protocol while operating on a emergency network. In addition, the content centric metadata information is also encapsulated into the bundle message as metadata information. In the DTN architecture, an application can send messages of arbitrary size called as Application Data Units (ADUs). The bundle layer transforms the ADUs into one or more Protocol Data Units (PDUs). The PDU is referred as \textit{bundle} which is forwarded by DTN nodes. In general, a bundle contains two or more \textit{blocks} of data in a defined format. Each block may contain either application data or other metadata used to deliver the bundle to its destination(s). At least two block structures are required to form a bundle: primary bundle block and payload block. Primary bundle block contains source/destination information on a bundle and expiration time (time-to-live). Bundle protocol supports the extension to the primary bundle block which allows specifying application specific metadata information. The CIDOR introduces the following metadata information for content-centric operation.

\begin{enumerate}
\item \textit{Nonce}: A randomly-generated byte string that is used to detect and discard duplicate \textit{Interest/Data} packet. Note that the Nonce is not the only way of detecting and discarding duplicates at a node. 
\item \textit{Bundle Expiration time (TTL)}: Lifetime of the bundle.
\item \textit{Bundle Type Extension block}: This field is used to define a packet as \textit{Interest} or \textit{Response}.
\item \textit{CIDOR PRIT Extension block}: This block contains a list of destination EIDs, which are interested for a particular \textit{Response}. 
\end{enumerate}

\subsection{Caching in CIDOR}
\label{oc}

The persistent storage of a DTN node stores the bundle until it gets the opportunity to forward the bundle to another node. The DTN node deletes the entry for a bundle for which the DTN node successfully forwards the bundle to the next opportunistic contact(s). Therefore, the persistent storage of the DTN node can not meet the future interests right after deleting the popular content. For this, the CIDOR allows the intermediate relay node to store the popular content for a while in the Opportunistic Content Store (LRU cache) to meet the future requests. Content providers store the content in the Content Store (CS). Nevertheless, each CIDOR searches a particular content in both the CS and the Opportunistic CS (if exist). In the disaster scenario, the opportunistic cache can enable the user to serve the critical information through the Opportunistic Content Store.

\subsection{CIDOR-PRIT structure for DTN}
\label{pit}

The CIDOR-PRIT table stores the requester EID of the \textit{Interest} packet. Upon reception of the \textit{Interest} packet, the CIDOR node checks its CIDOR-PRIT table. If there is a match, it adds the requester EID to that entry and drops the packet. While handling the response packet, the CIDOR node checks its CIDOR-PRIT table to find if there is any pending requester(s) for this response. If there is a match, the node adds the pending requester EID(s) in the Bundle PIT Extension block of the response packet.

\begin{figure*}[!t]
\centering

\subfloat[A, B, C are relaying similar interests from multiple requesters]{ 
\centering
\includegraphics[width=.8\columnwidth]{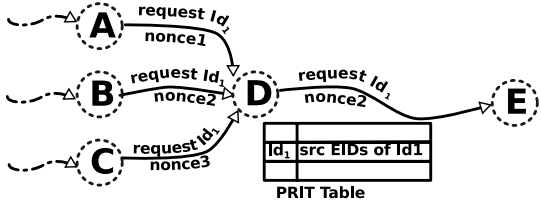}%
\label{dtn1}}
\hfill \centering
\subfloat[A, B relaying similar interests of same requester whereas C is relaying similar interest of different requester. ]{\includegraphics[width=.8\columnwidth]{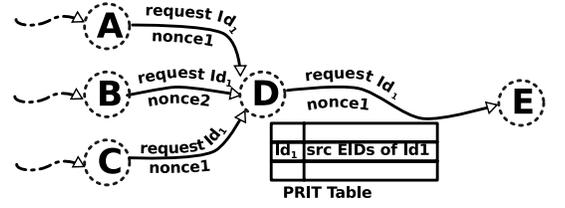}%
\label{dtn2}}

\subfloat[A is relaying response for request id1 that is heading towards S.]{\includegraphics[scale=0.45]{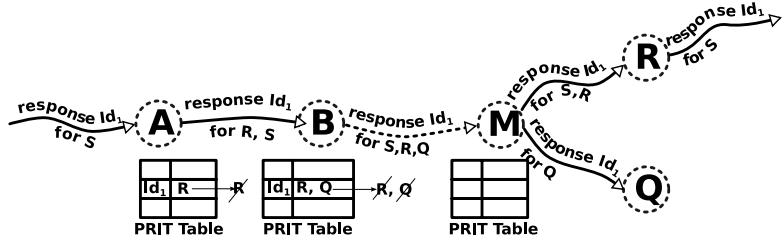}%
\label{response}}

\hfill
\caption{A simple scenario of duplicate suppression and redundancy elimination of the CIDOR (a) interest forwarding policy (similar requests from multiple requesters), (b) duplicate reduction; requests containing the same nonce value are suppressed (c) Response forwarding policy.}
\label{dup}
\end{figure*}

\begin{algorithm}
\algsetup{linenosize=\small}
\footnotesize
\caption{Handling Interest Message}
\label{interest}
\begin{algorithmic}[1]

\STATE $key \leftarrow [Interest]$
\IF{$key\ in\ Local\ Cache $}
   \STATE $content \leftarrow Cache(key)$ 
 \ENDIF 
 
\IF {$content \neq NULL$}
 \STATE $sourceEID \leftarrow [Interest] $
 \STATE $response \leftarrow createResponse(content)$
 \STATE $addPRITExtBlock(response)$
 \STATE $Send\ response\ back\ to\ sourceEID$
\ELSE 
   \IF{$ myEID $ = $destinationEID$}
     \IF {$TTL\ not\ expired $}
     \STATE $destinationEID \leftarrow newRandomHost$
     \ENDIF
   \ENDIF

\STATE $Add\ source\ EID\ to\ PRIT\ table$
\STATE $forward\ the\ Interest\ to\ next\ Hop $

\ENDIF

\end{algorithmic}
\end{algorithm}

\subsection{Interest Propagation}
 
On the reception of an \textit{Interest} packet, a node applies the handling \textit{Interest} packet algorithm as shown in \textbf{Algorithm~\ref{interest}}. First, it searches its opportunistic cache. If no match is found, it searches its \textit{Content Store}. If the node does not find a match in its \textit{Content Store}, it adds the \textit{source EID} of the \textit{Interest} packet into the CIDOR-PRIT table. These \textit{source EID}(s) are used as the \textit{destination EID}(s) in the response. The duration of a request propagation solely depends on the bundle expiration time (TTL). When a bundle expires, the BPA deletes the bundle.

\begin{algorithm}
\algsetup{linenosize=\small}
\small
\caption{Handling Response Packet}
\label{responsealg}
\begin{algorithmic}[1]
\IF {$myEID$= $destinationEID $}
   \STATE $notify\ application$
   \IF {$ PRITBlock\ is\ empty $}
     \STATE $drop\ the\ packet$
     \RETURN
	\ENDIF  
 \ENDIF  

 \IF {$myEID \in PRITBlock $}
    \STATE $notify\ application$
    \STATE $ PRITBlock \leftarrow PRITBlock \setminus \{myEID\}$   
 \ELSE 
     \STATE $key \leftarrow [Response]$
	 \STATE$record \leftarrow PRITtable(key)$
     \IF {$record \neq NULL$}
		\STATE $srcEID \leftarrow getRequester(record) $
		\STATE $PRITBlock \leftarrow \cup \{srcEID\} $
      \ENDIF
\ENDIF 
\STATE $add\ content\ to\ opportunitistic\ cache$
\STATE $Forward\ response\ to\ next\ hop$

\end{algorithmic}
\end{algorithm}

\subsection{Response Forwarding}

When the \textit{Interest} packet reaches a node having content in its content store, the node constructs a \textit{response bundle} with the content and sends it back to the originator of the request. The content provider inserts the \textit{bundle expiration time} from the received \textit{Interest} packet. On the reception of the \textit{response bundle}, a node follows the handling response message algorithm as shown in \textbf{Algorithm~\ref{responsealg}}. Intermediate nodes check the CIDOR-PRIT table and removes the entry if there is a match for the response bundle. If a match is found in the PIT table, the node adds all the \textit{Source EIDs} of originator(s) in the \textit{bundle extension block}. When the \textit{response bundle} reaches the originator of the \textit{Interest} packet, it checks the \textit{bundle extension block}. If the extension block is not empty, it updates the destination host list of the \textit{response bundle} by deleting itself from the list and waits for the next opportunistic contact.

\subsection{Duplicate Suppression}
\label{duplicate}

If multi-copy DTN routing strategies are used, both \textit{Interest and Data} packets get duplicated and are spread into the network. This duplication is inherent in the DTN routing which can also suppress some duplicates in every node. However, messages in the DTN routing are generally distinguished by the tuple of the source and destination addresses and a unique identifier assigned by the source end-point of these messages. The host-centric DTN has no way of detecting loops at the content level. Nevertheless, our proposed architecture includes some distinct parameters (e.g., \textit{Nonce}) for detecting and preventing such duplication. 

For instance, Figure~\ref{dtn1} and~\ref{dtn2} illustrates the duplicate suppression in an intermediate node. In Figure~\ref{dtn1}, node A, B, and C are relaying the similar interests of different nonce values. The similar interest of different nonce values indicates the similar requests from different requesters. Upon reception of these interests, D keeps records of requester information on the \textit{Interest} packet in the CIDOR-PRIT table and forward only one \textit{Interest} packet to the next opportunistic contact. In Figure~\ref{dtn2}, D receives the similar interests from the intermediate relay nodes (A and C) containing the same nonce value. Therefore, D keeps record of one \textit{Interest} packet in the CIDOR-PRIT table and discard the other one. CIDOR node also maintains a data structure referred as \texttt{processedMessageList}. This structure keeps records of the seen message (\textit{Interest/Data}) packets for a while, so that intermediate nodes can detect the forwarding loop and drop the same message.

\subsection{Redundancy Elimination}
\label{re}

The CIDOR can reduce the generation of redundant \textit{Interest/Data} packets by maintaining the CIDOR-PRIT table. Figure~\ref{dup} illustrates the scenario of reducing the forwarding of redundant \textit{Interest} packets. As discussed in Section~\ref{duplicate}, the intermediate CIDOR nodes aggregate the similar interests in the CIDOR-PRIT table. The \textit{Interest} packets containing a different nonce value are recorded in the CIDOR-PRIT table. CIDOR-PRIT table is exploited while forwarding the response. Each response packet contains a \textit{CIDOR PRIT extension block} which contains a list of the pending requester(s) information provided by the CIDOR-PRIT table. Upon the reception of a response packet, each node checks its CIDOR-PRIT table to verify the pending requester(s) for this response. For example, in Figure~\ref{response}, node A receives a response heading to the node S. A checks its PRIT table and finds a pending requester R for this response. Node A, therefore, adds R into the PRIT extension block of the response packet. Similarly, node B adds Q in the block. Subsequently, the node M meets the node R and Q. Then, the node M creates two copies of this response and forward to R and Q. The node R keeps its copy and forward another copy towards S.

\subsection{Compatibility Consideration}

The CIDOR has a good compatibility with the native DTN architecture due to the following design principles. First content-centric functionalities remain in the CIDOR control plane decision engine. CIDOR does not modify the native BP. The \textit{Interest/Data} packet is encapsulated in the BP packet in addition to the content-centric meta-information. The CIDOR exploits the BP extension blocks to specify content-centric meta-information. Second not all the nodes are required to implement CIDOR since the nodes (vanilla DTN) perform as the intermediate relay nodes forwarding the packet to the next opportunistic contact. Finally CIDOR separates the forwarding plane from the routing and therefore, the existing DTN routing or the content-based DTN routing can easily be adapted to the CIDOR as shown in Figure~\ref{CIDOR_node}.

%% file: performance.tex
\section{\textbf{Performance Evaluation}}
\label{result}

We evaluate the prototype of the CIDOR architecture using the Opportunistic Network Simulator (ONE) \cite{keranen2009one}. The primary goal of our evaluation is investigating the efficiency of our proposed architecture in terms of the response ratio and latency with respect to the availability of resources in the network. Besides this, we also investigate other evaluation metrics (see Section~\ref{metric}). For this evaluation, we run the simulation with different classes of DTN routing strategies in different mobility scenarios. Our experiment is divided into 4 phases. \textbf{(\rom{1})} We run simulation in the Helsinki city scenario (Section~\ref{city}) by varying the buffer size and fixed TTL 500s. From this phase, we choose a suitable value of the buffer size, at which all routing achieve a good performance in terms of the response ratio. \textbf{(\rom{2})} We fix the buffer size from the phase \rom{1} and evaluate the effect of TTL on the performance. Then we choose a suitable TTL value at which all routing (see Section~\ref{dr}) achieves a good performance. \textbf{(\rom{3})} With the fixed buffer size and the fixed TTL value (from phase \rom{1} and \rom{2}), our experiment follows by varying the number of producers. \textbf{(\rom{4})} With the fixed buffer size, TTL and number of producers, we continue our experiment using Random Way Point (RWP) mobility model by varying the number of resources of each producer. We use both uniform distribution and Zipf distribution to generate queries in the simulation. We plot the average result of 10 simulation runs for each figure.

\begin{figure*}[!t]
\centering

\subfloat[Uniform]{\includegraphics[width=1.5in]{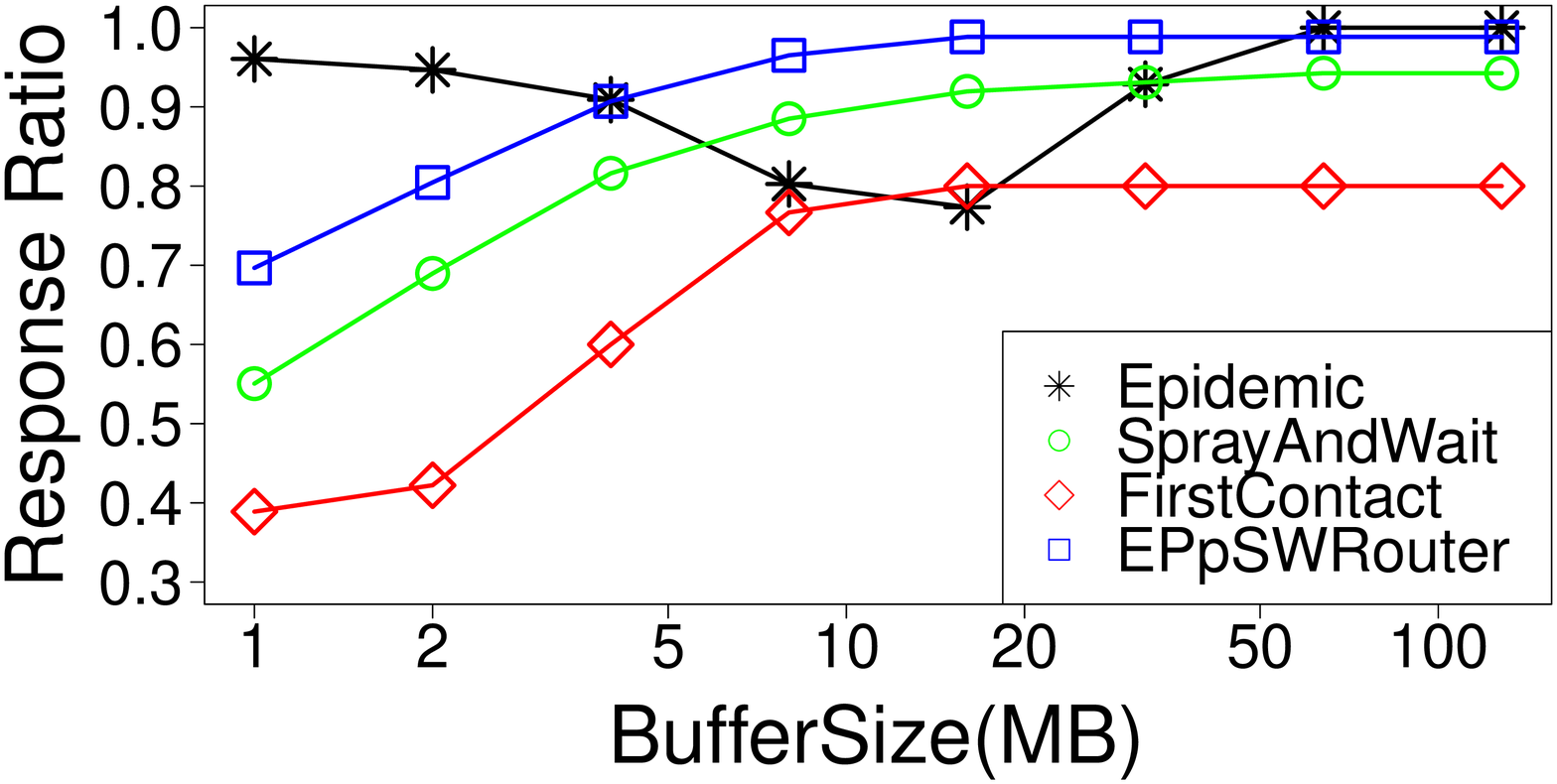}%
\label{a}}
\hfil
\subfloat[Zipf]{\includegraphics[width=1.5in]{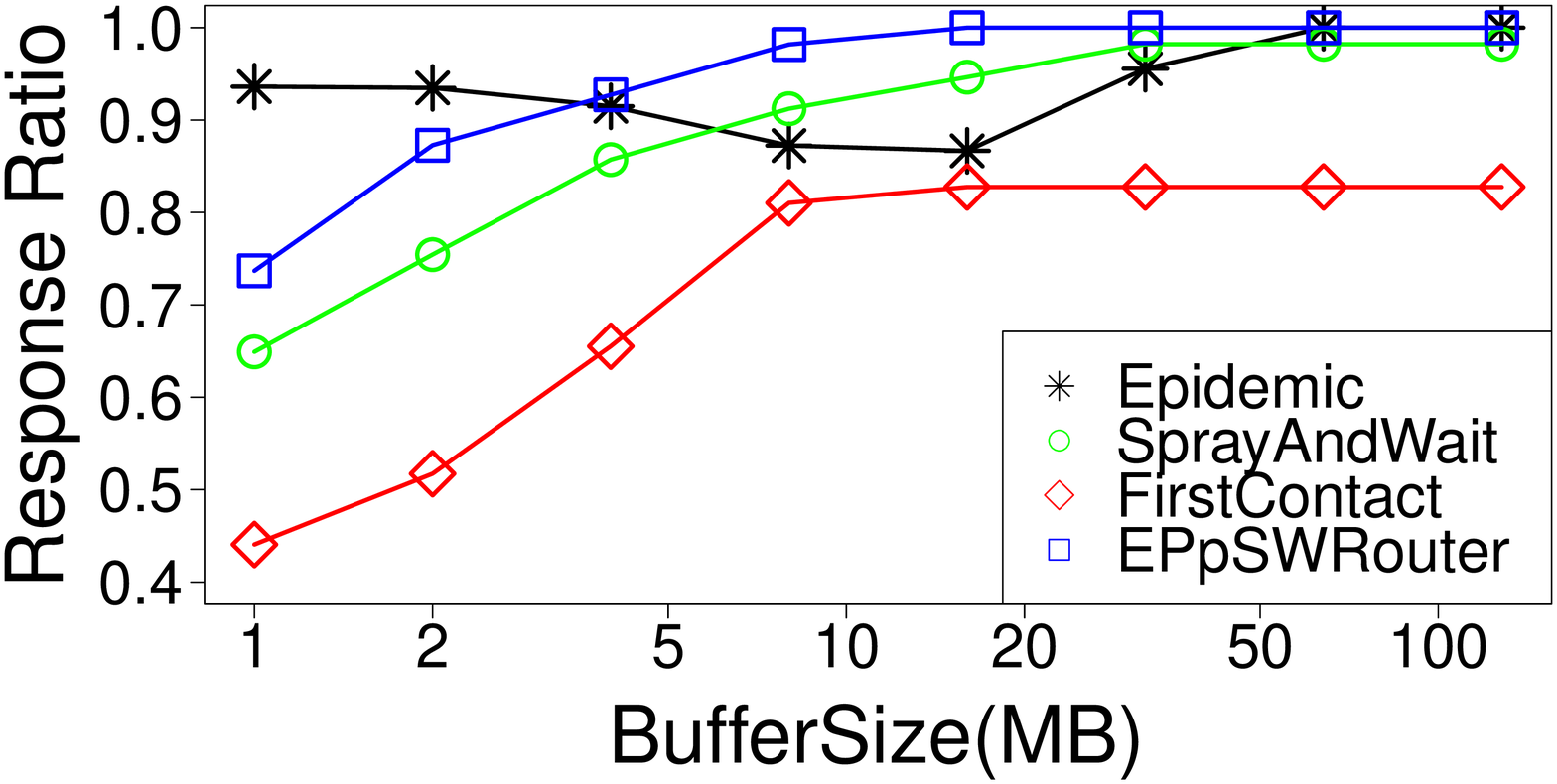}%
\label{b}}
\hfil
\subfloat[Uniform]{\includegraphics[width=1.5in]{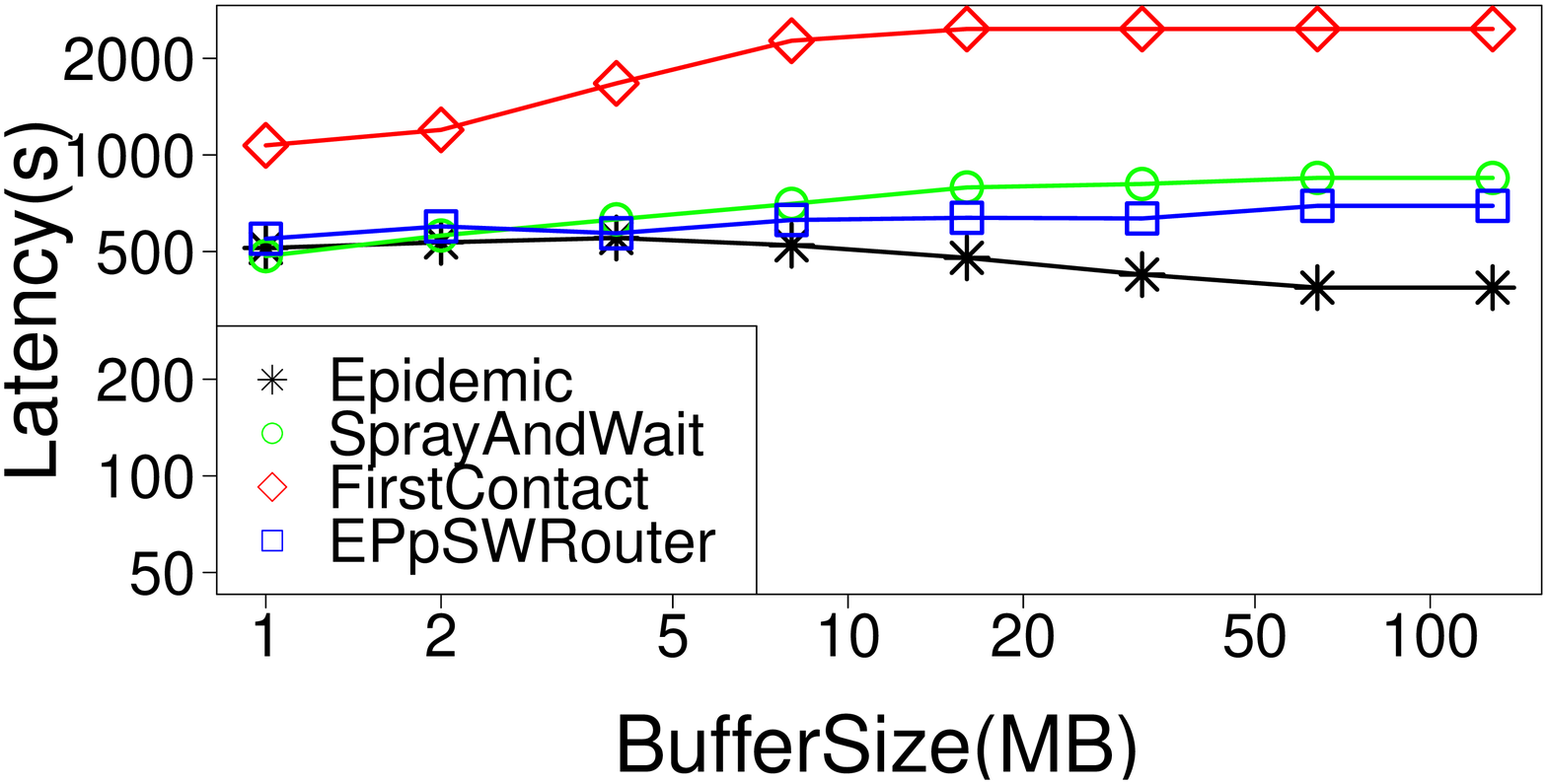}%
\label{c}}
\hfil
\subfloat[Zipf]{\includegraphics[width=1.5in]{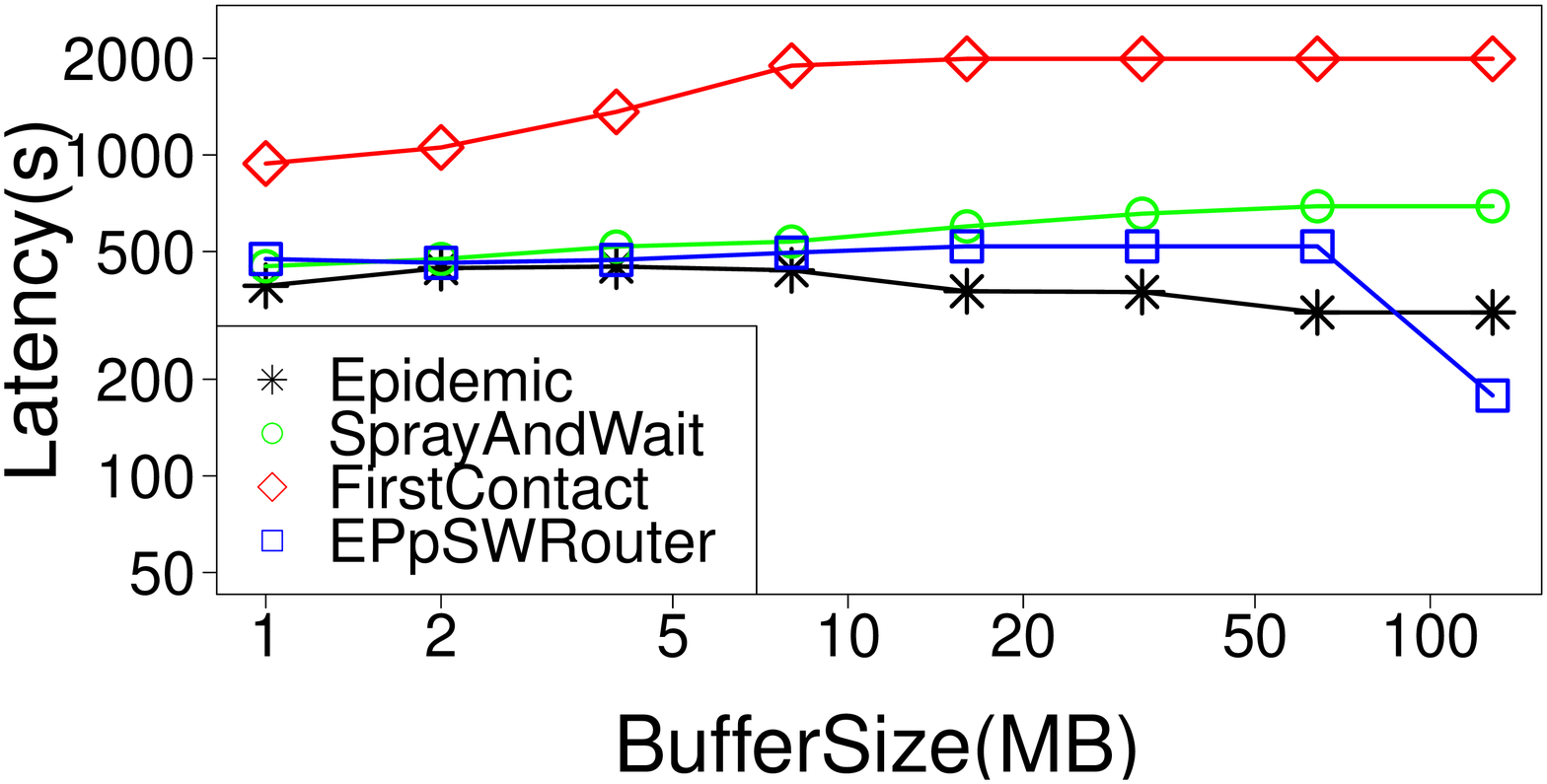}%
\label{d}}

\vspace{.2cm}

\subfloat[Uniform]{\includegraphics[width=1.5in]{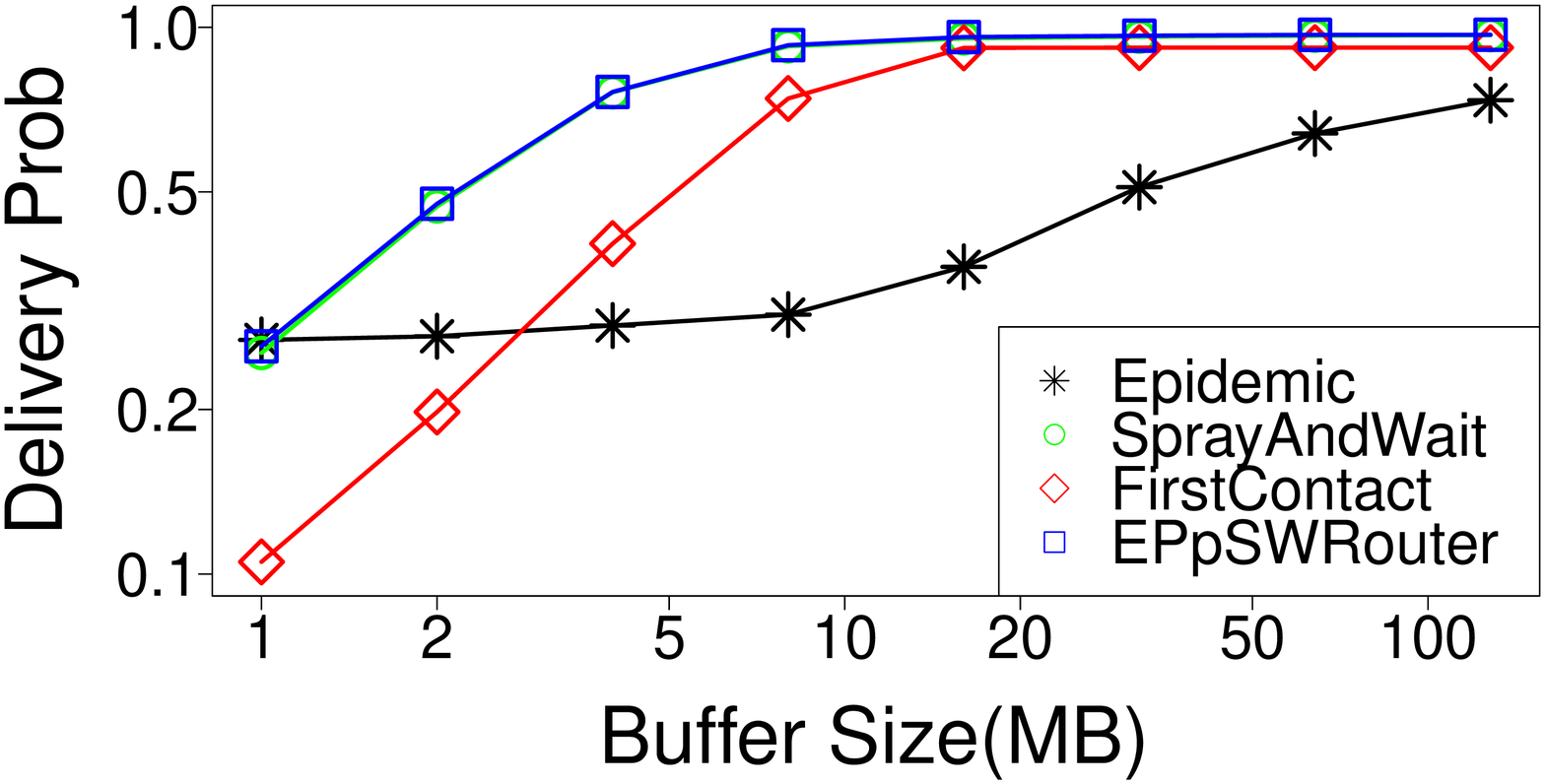}%
\label{e}}
\hfil
\subfloat[Zipf]{\includegraphics[width=1.5in]{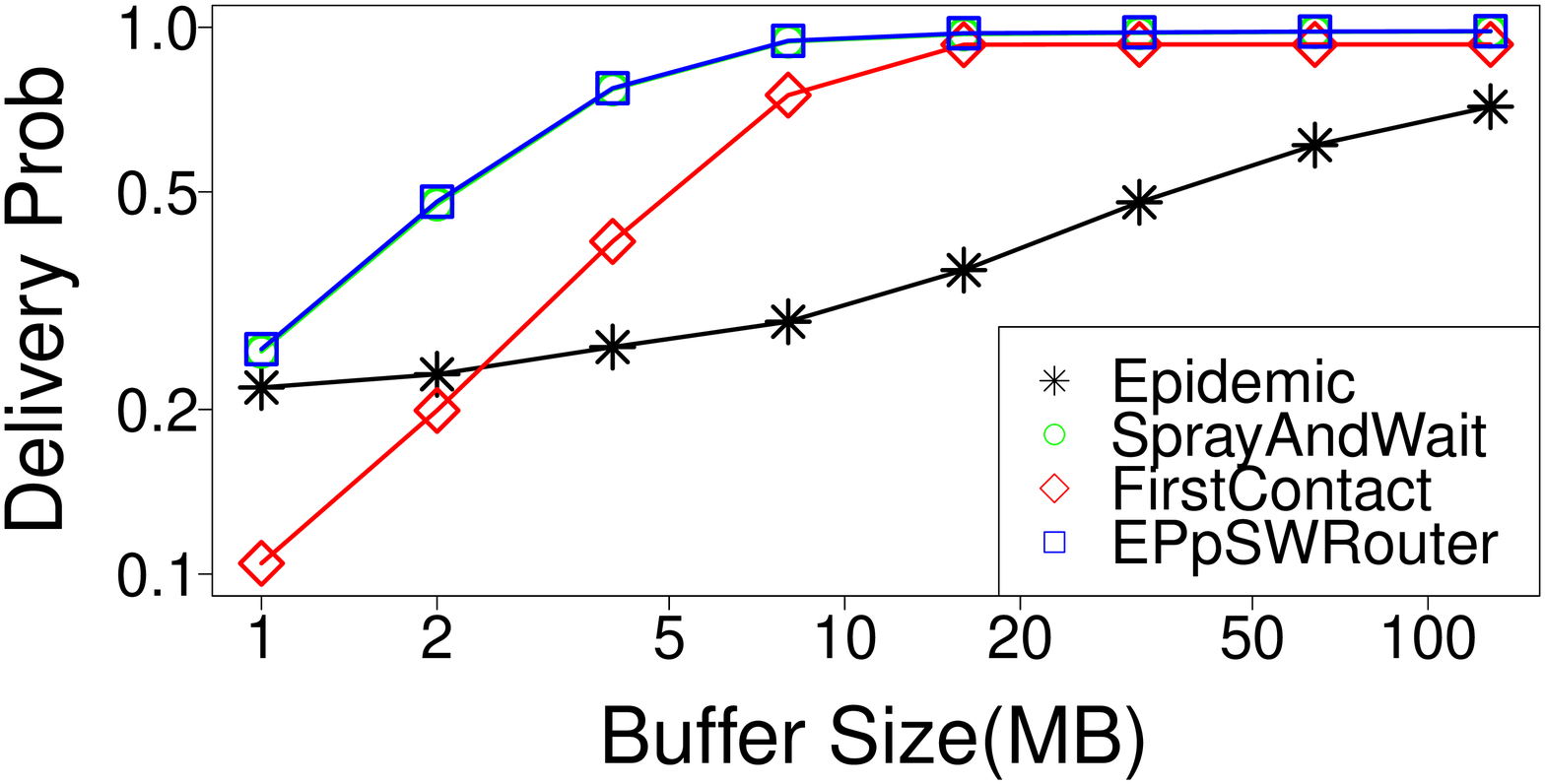}%
\label{f}}
\hfil
\subfloat[Uniform]{\includegraphics[width=1.5in]{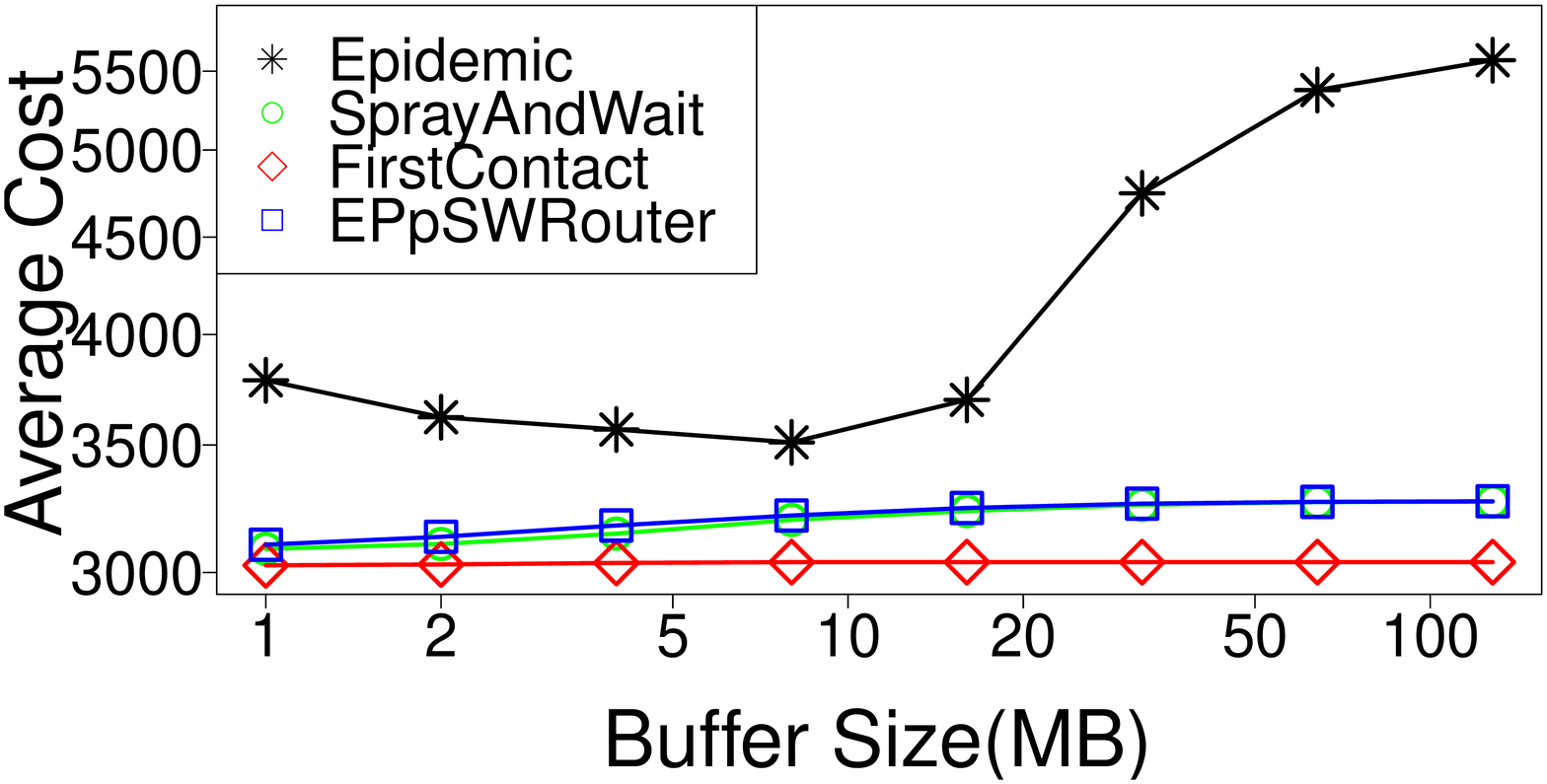}%
\label{g}}
\hfil
\subfloat[Zipf]{\includegraphics[width=1.5in]{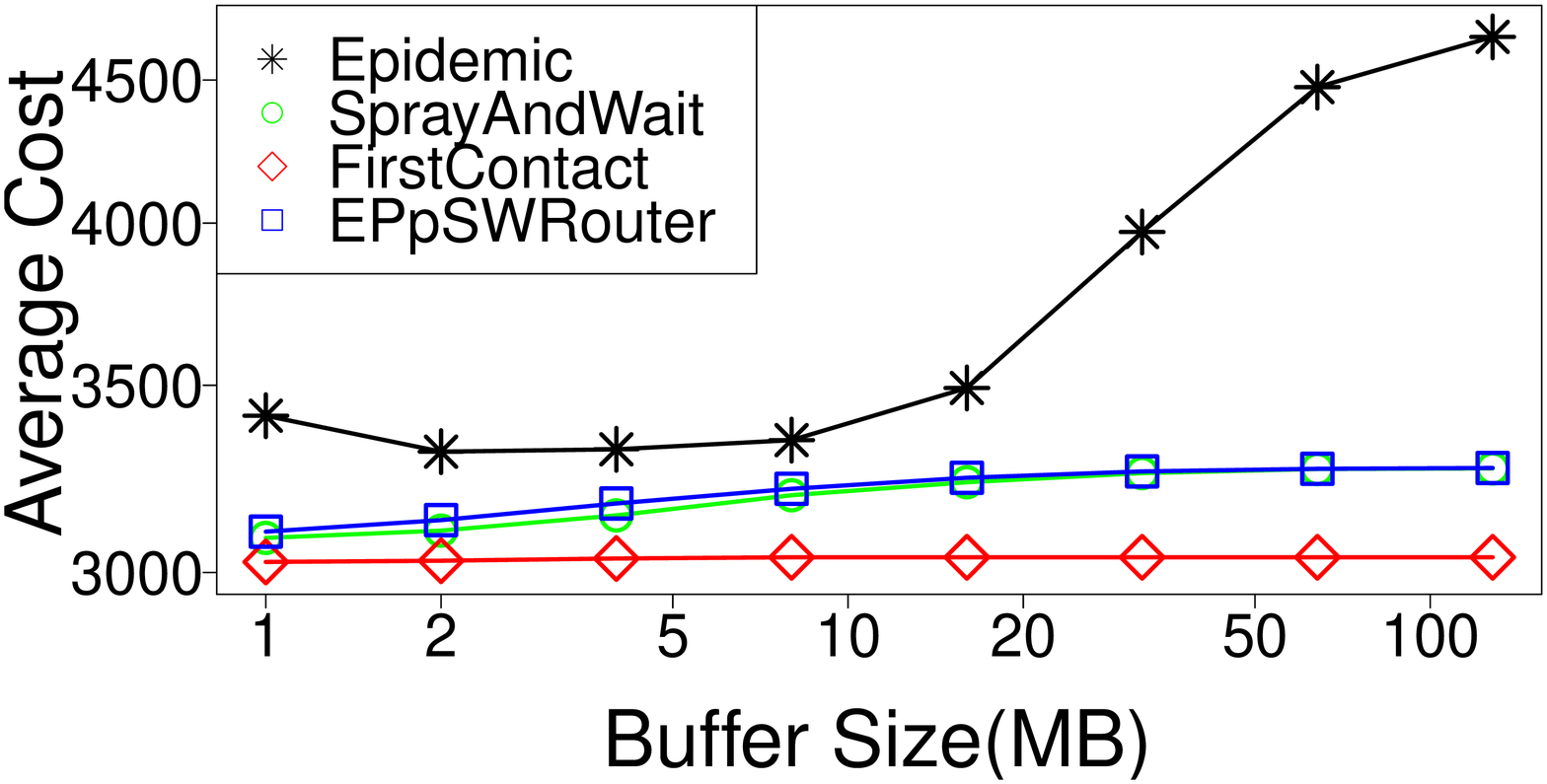}%
\label{h}}

\caption{The behaviour of CIDOR with different DTN routing varying the buffer size in Helsinki City Scenario.}
\label{helsinki}
\end{figure*}

\subsection{Helsinki City Scenario} 
\label{city}

The ONE simulator provides the map of Helsinki city area (e.g., roads and pedestrian walkways). Besides, three Map-based Movement models are incorporated in the simulator: 1) Random Map-Based Movement, 2) Shortest Path Map-Based Movement, and 3) Routed Map-Based Movement. In Random Map-Based Movement model, nodes move randomly following the paths provided by the map data. However, the random walk may not be a very accurate approximation of real human mobility. In contrast, Shortest Path Map-Based Movement is a more pragmatic model where the nodes select a destination point randomly from a list of Points of Interest (POI) on the map and choose the shortest path to that point. The list of POI may include the popular destinations (e.g., shops, restaurant, tourist attractions). Nevertheless, our simulation uses the Shortest Path Map-Based Movement model. Routed Map-Based Movement model considers the pre-determined routes (e.g., bus, tram or train routes) of the nodes.

\subsection{DTN Routing}
\label{dr}

The CIDOR operates independently of the routing strategies. The forwarding function determines which exit interface to use to send the packet to next hop.  For our experiment, we choose three different routing strategies: \textit{Epidemic} \cite{vahdat2000epidemic}, \textit{Spray-and-Wait} \cite{spyropoulos2005spray}, \textit{First contact} \cite{jain2004routing}. The first two represent multi-copy routing scheme, whereas the third one represents single-copy routing scheme. \textit{Epidemic} routing has no limitation of generating copies of each message. Each node of \textit{Epidemic} router carries a list of all pending messages to be delivered. Subsequently, the node exchanges all messages to the next opportunistic contact that are not common on their list. \textit{Spray-and-Wait} generates a limited number of copies for every message and spreads initially. If a node does not find the destination in spray phase, it waits for the destination to perform direct transmission. In our experiment, \textit{Spray-and-Wait} generates 10 copies for every message in the spray phase while \textit{First contact} generates only one copy per message. In addition, we experiment the CIDOR with a combination of the \textit{Epidemic} and the \textit{Spray-and-Wait} (referred as \textit{EP$_{p}$SWRouting}). \textit{EP$_{p}$SWRouting} broadcasts \textit{Interest} packets using the Epidemic (limited by 10 copies) to reach the potential \textit{content providers} faster. When the \textit{Interest} packet reaches the content source, \textit{EP$_{p}$SWRouting} uses the \textit{Spray-and-Wait} routing to deliver the content back to the requester.

\subsection{Query Distribution}
\label{dist}
The simulation generates user requests based on a query key range Q with size N = 100 assuming that query key Q$_{j} \in $K is the $j^{th}$ popular content in the network. The probability of each query Q$_{j}$ issued by an user is randomly  selected from a normal distribution with P$_{j}$ as the mean value. In real applications, the popularity of content is correlated with user requests \cite{liu2005client} and follows the well-known Zipf distribution \cite{breslau1999web}. Therefore, we exploit the Zipf distribution and uniform distribution for generating P$_{j}$ of different query keys. All content items are enumerated in the content ladder and assigned a probability of appearance P$_{j}$ for the Zipf distribution with parameter 1 and normalizing constant 0.2. A number of distinct resources matching with each query is distributed in the network. Let us assume that the list of popular contents is [A, B, C, D]. The probability of appearance, then, is [0.2, 0.1, 0.067, 0.05]. Assuming the value range (0, 1000), the content ladder contains [200 (0.2$\times$1000), 300 ((0.2+0.1)$\times$1000), 367 ((0.2+0.1+0.067)$\times$1000), 417]. While generating requests, the requester picks up a random value from the range (0,1000). For instance, if the random number is 333, then the requester checks the content ladder to determine the position of 333 (index 2). The corresponding content of index 2 from the list of popular content is C. 


\subsection{Evaluation Metrics}
\label{metric}

\begin{itemize}
\item Response Ratio. The probability for retrieving content in response to an \textit{Interest} packet issued by a node. 
\item Latency. The average amount of time passed to receive content in response to a request.
\item Delivery Ratio. The average ratio of the total number of successfully delivered messages with respect to the total number of messages sent.
\item Average cost. The average number of content transmissions required to deliver a data item. It also includes all duplicate messages during transmissions. 
\end{itemize}

\begin{figure*}[!t]
\centering

\subfloat[Uniform]{\includegraphics[width=1.5in]{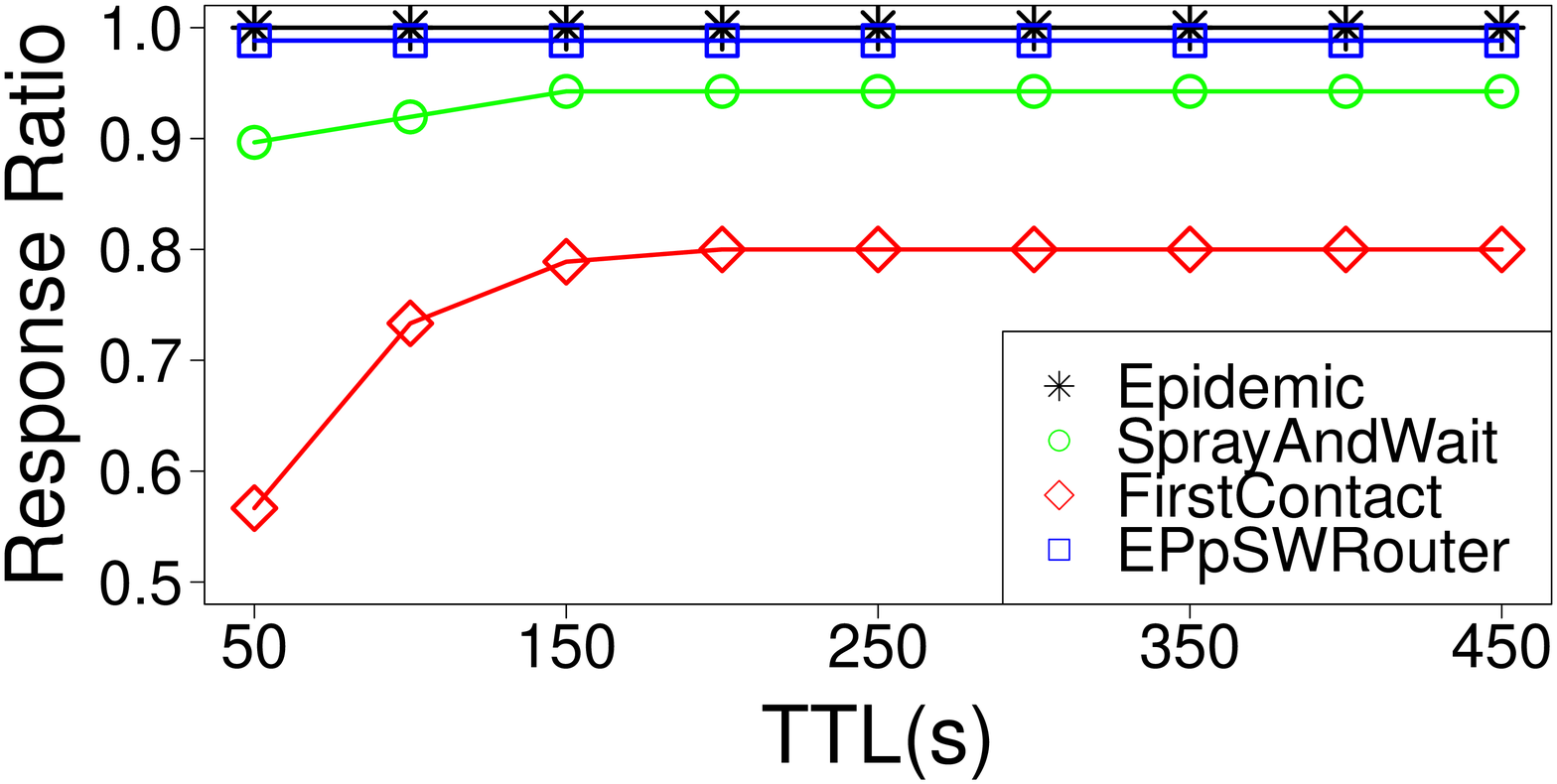}%
\label{i}}
\hfil
\subfloat[Zipf]{\includegraphics[width=1.5in]{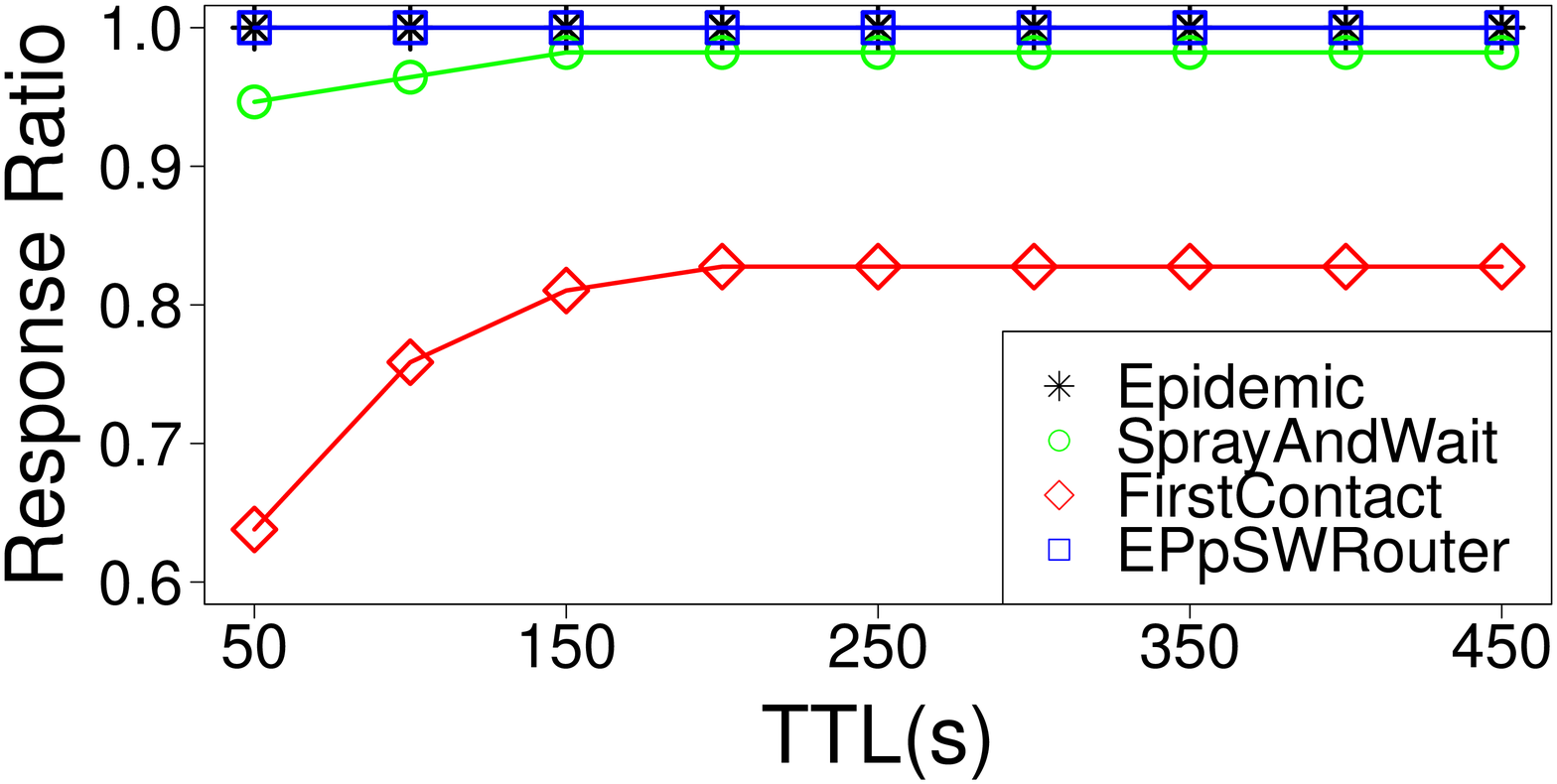}%
\label{j}}
\hfil
\subfloat[Uniform]{\includegraphics[width=1.5in]{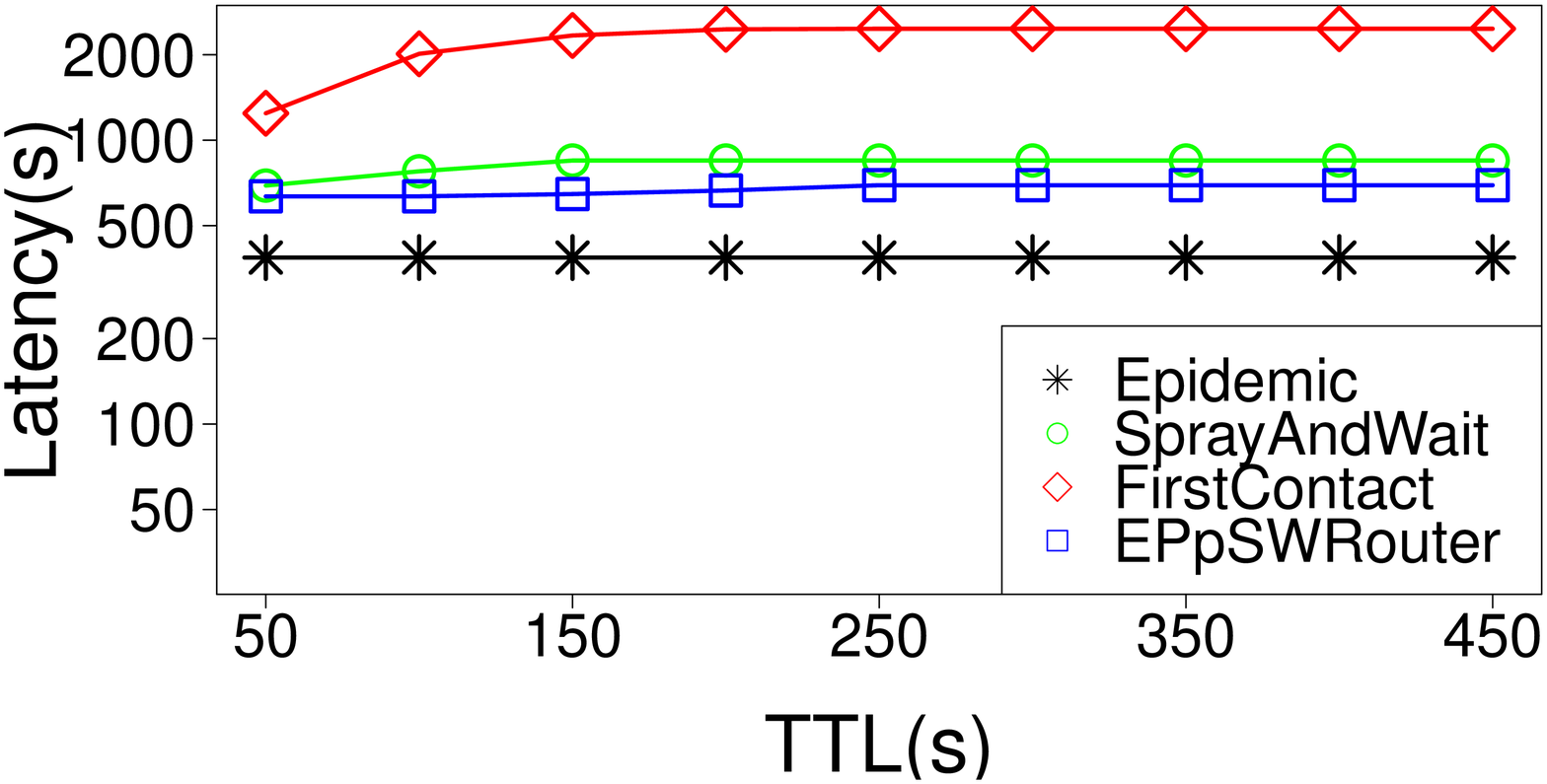}%
\label{k}}
\hfil
\subfloat[Zipf]{\includegraphics[width=1.5in]{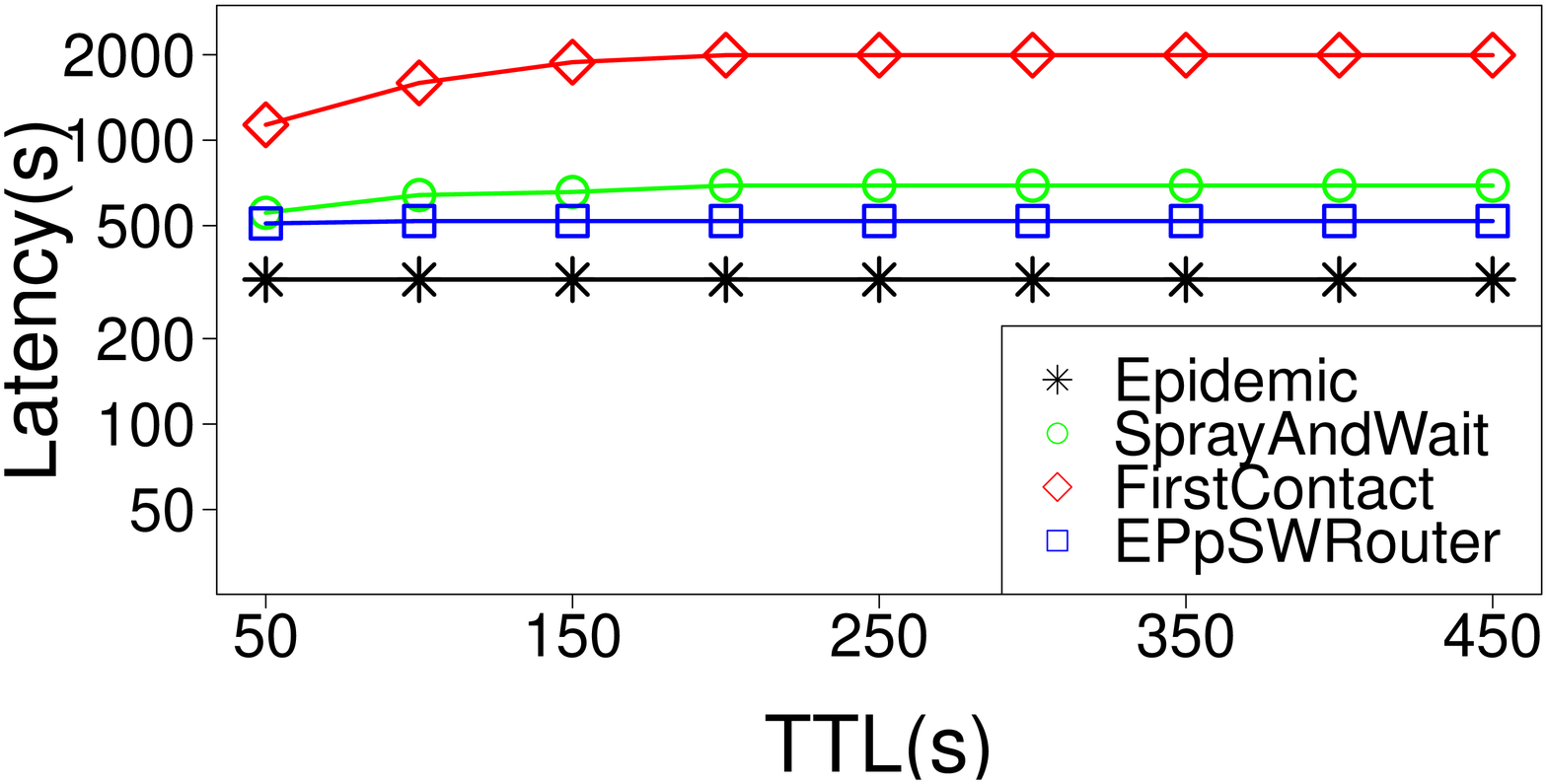}%
\label{l}}

\vspace{.2cm}

\subfloat[Uniform]{\includegraphics[width=1.5in]{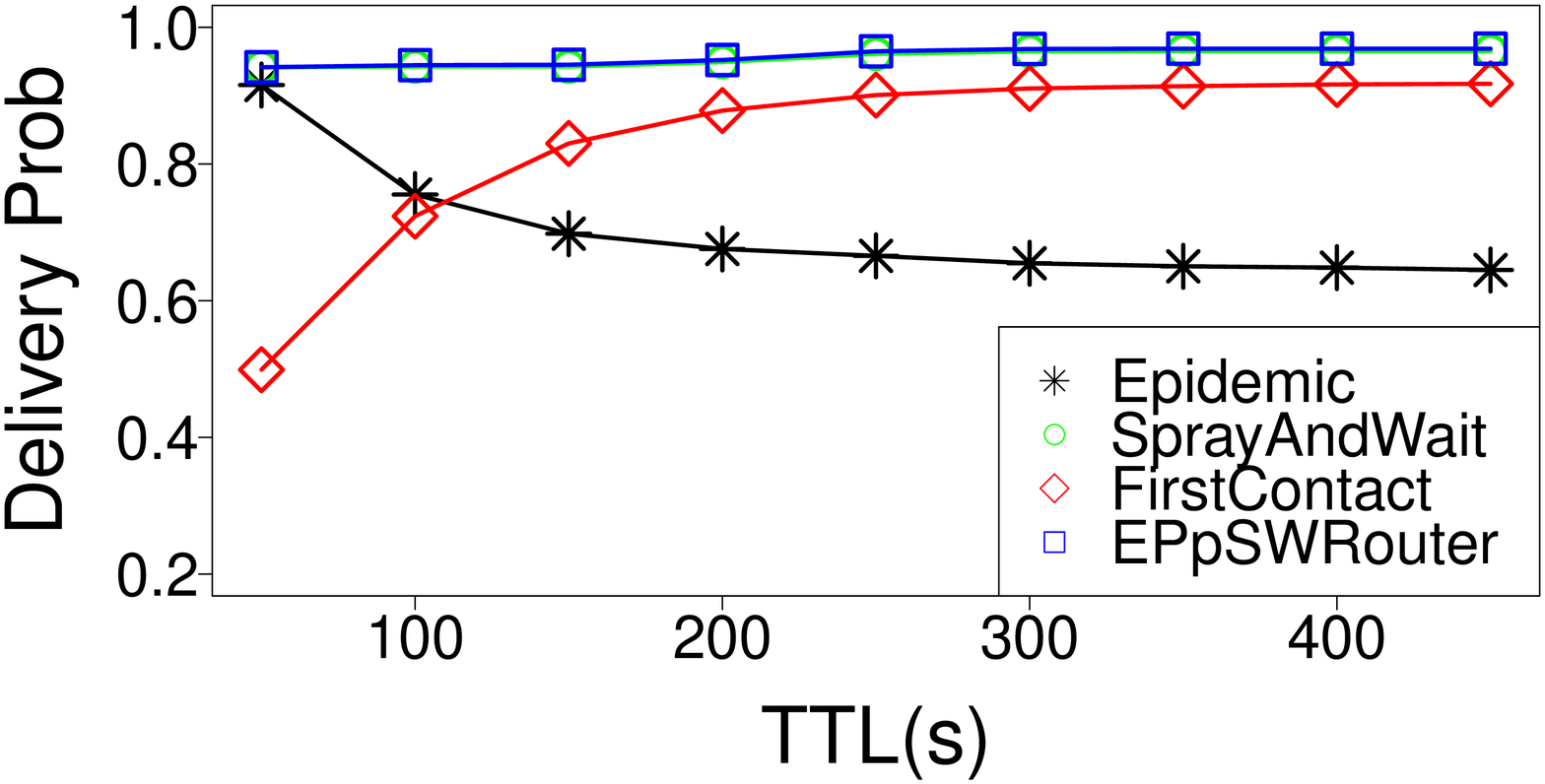}%
\label{m}}
\hfil
\subfloat[Zipf]{\includegraphics[width=1.5in]{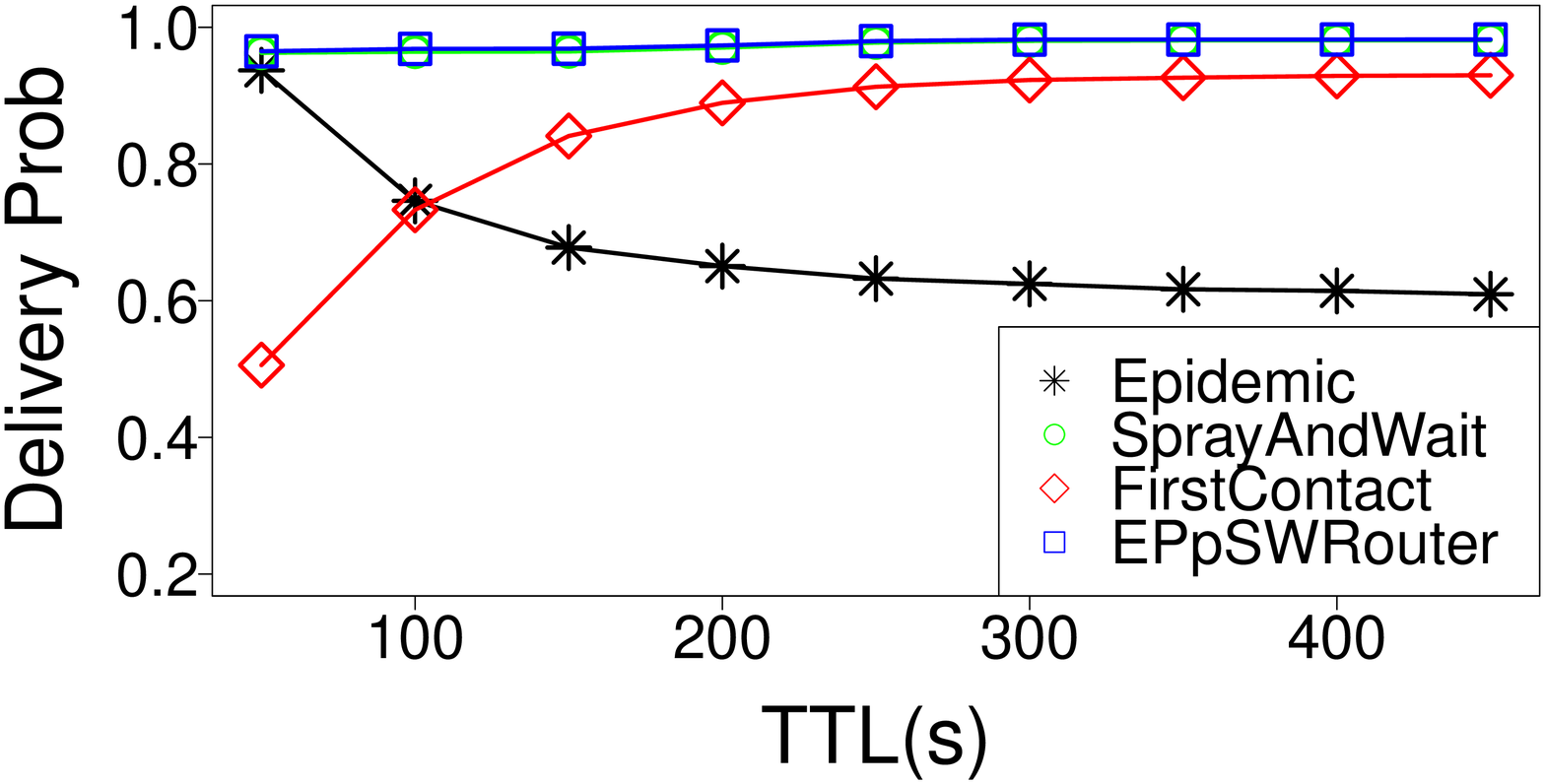}%
\label{n}}
\hfil
\subfloat[Uniform]{\includegraphics[width=1.5in]{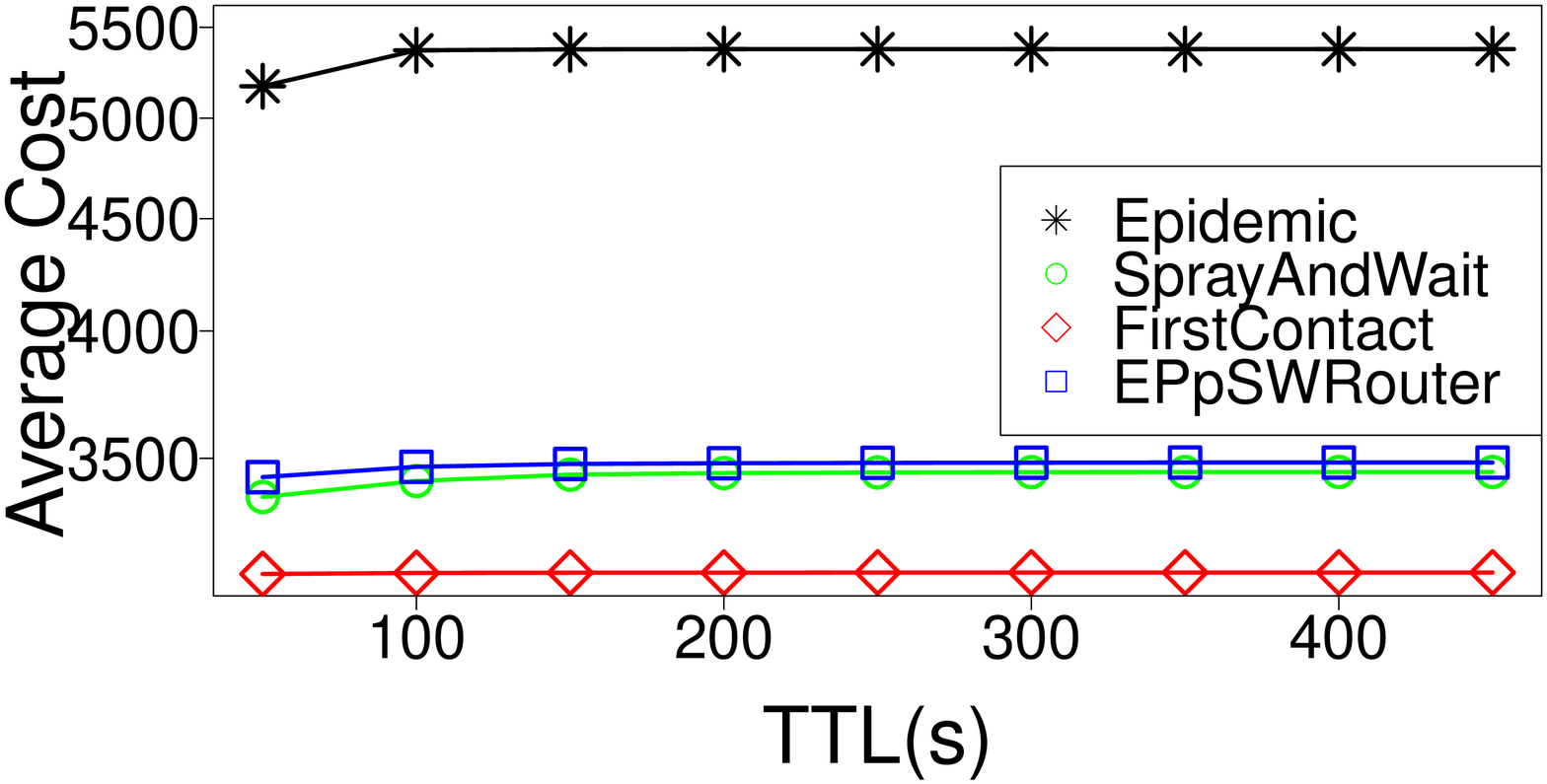}%
\label{o}}
\hfil
\subfloat[Zipf]{\includegraphics[width=1.5in]{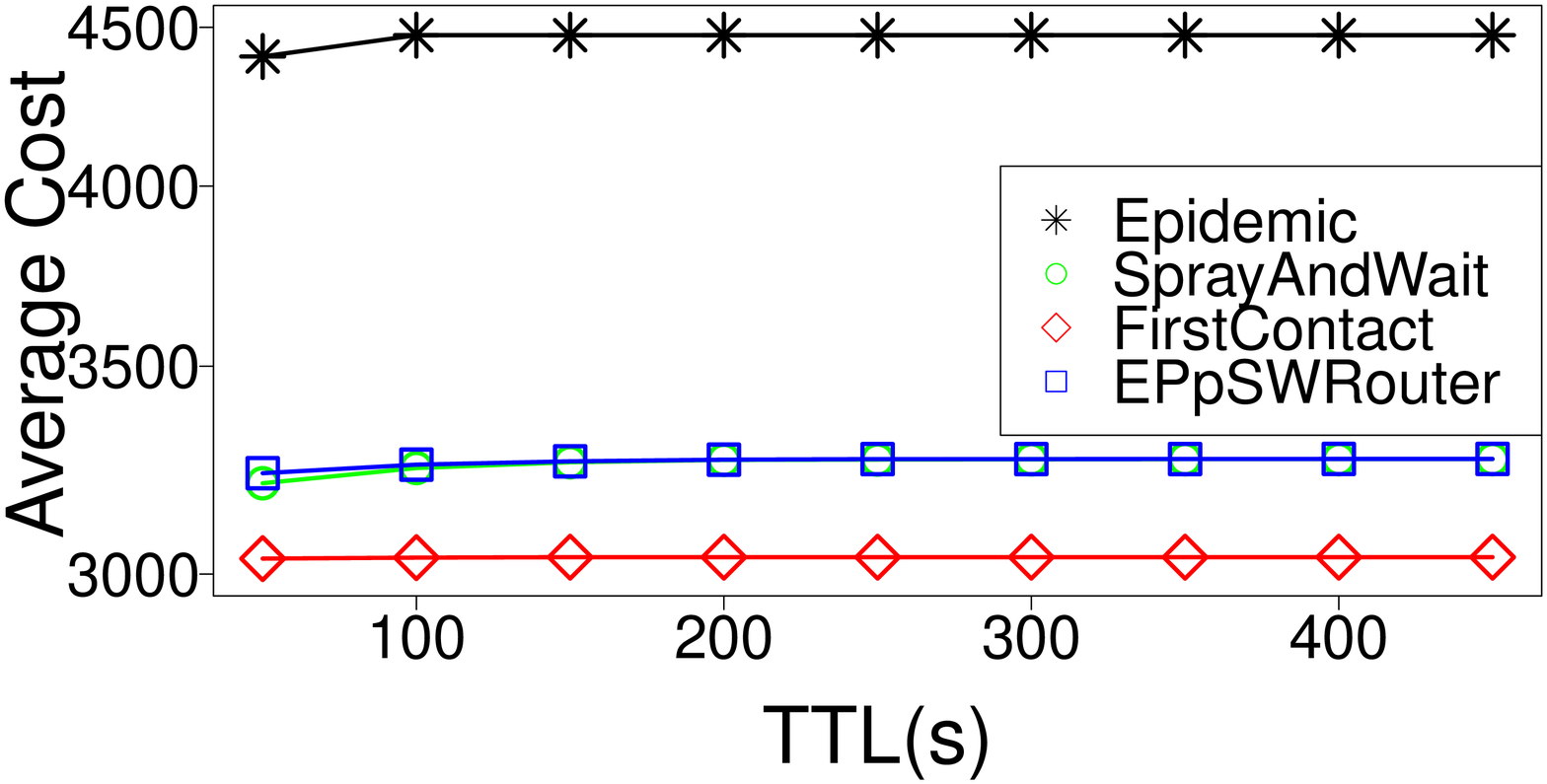}%
\label{p}}

\caption{The behaviour of CIDOR with different DTN routing strategies varying the TTL in Helsinki City Scenario.}
\label{helsinki1}
\end{figure*}

\subsection{Experimental Settings}

Our experiment uses two different mobility models: Map-based mobility model and well known Random Way Point (RWP). Each model contains three groups of users: \textit{content provider}, \textit{requester}, and intermediate nodes. The Map based movement model constraints the node movements to actual streets provided by the map of the Helsinki city. Users move with realistic speed along the shortest paths between different points of interests (POIs) and random locations. Nodes are divided into four different groups and assigned with different probabilities to choose the next group specific POI or random places to visit. The simulation area approximately is 4.5km x 4.5km. For Helsinki scenario, we use 5 requesters, 35 intermediate nodes, and 10 content providers. A brief overview of all the simulation parameters is presented in Table~\ref{simulation}. We evaluate our metrics by varying the number of resources in the RWP mobility scenario. For this simulation, nodes move with relatively slower speed than vehicles in the range of 0.5-1.5 m/s.

\begin{table}
\small
\centering
\caption{Simulation Parameters}
\label{simulation}
\begin{tabular}{ |l|l| } \hline  
Parameter & Value \\ \hline 
Simulation Duration & 86400s \\ 
Number of Requesters & 5 \\
Time interval of generating Interests & 100s \\ 
Number of Intermediate nodes & 35 \\
Number of Producers & 10:[1-15] \\
Buffer Size & 64M [1M-128M] \\
TTL value & 500s:[50-500s] \\
Transmission range & 100m \\ 
Transmission speed & 2500Kbps. \\ \hline
\end{tabular}

\label{SimulationParameters}
\end{table}

\subsection{Effect of Buffer Size}

In this experiment, we vary the buffer size of each node and evaluate the performance of the CIDOR with four different DTN routing strategies. In general, a larger buffer size helps achieving a higher delivery and response ratio. As expected, the \textit{Epidemic} routing achieves a higher response ratio and a reduced latency with a larger average cost. The response ratio of the \textit{Epidemic} routing tends to increase when the buffer size is more than 16MB (Fig~\ref{a},~\ref{b}). We observe that the response ratio of the \textit{Epidemic} routing decreases gradually up to the buffer size 16MB. This is because the router does not have sufficient space to accept all the requests due to the lower buffer size. Fig~\ref{e} and~\ref{f} show that the delivery ratio is quite low, around 30$\%$. However, the delivery ratio and the average cost start increasing when the buffer size is more that 16MB (Fig~\ref{g},~\ref{h}). In addition, the response ratio tends to increase with the increase of the buffer size from 16MB. The average latency to retrieve the contents is reported in Fig~\ref{g} and~\ref{h}. Among all routings (\textit{First contact, \textit{Spray-and-Wait}, \textit{Epidemic}, \textit{EP$_{p}$SWRouting}}), the \textit{First contact} shows a poor performance in terms of the response ratio and latency, because the \textit{First contact} generates a single copy for each request and therefore, the average cost is lower compared to other routings. In contrast, when the buffer size increases, the \textit{EP$_{p}$SWRouting} achieves a higher response ratio with a slightly higher latency than \textit{Epidemic} routing. Nevertheless, if the buffer size is sufficiently large (e.g., 64MB in our settings), the response ratio gets saturated for all the routing schemes.

The notable finding is that the delivery ratio of the single copy (\textit{First contact}) and multi-copy (\textit{Spray-and-Wait} and \textit{EP$_{p}$SWRouting}) routing scheme is much higher than \textit{Epidemic} routing, while achieving a comparable response ratio. This is because the CIDOR applies duplicate suppression mechanism to suppress the duplicates.

\subsection{Effect of TTL}

A greater value of TTL increases the probability of a request reaching the content provider. With a large TTL, nodes get more time to relay the messages towards the potential destinations, therefore the delivery ratio of all DTN routings increase accordingly (Fig~\ref{m},~\ref{n}). Fig~\ref{i},~\ref{j},~\ref{k}, and~\ref{l} show that the \textit{Epidemic} outperforms all the other routings in terms of the response ratio and latency. However, the average cost of the \textit{Epidemic} routing is considerably larger than all of the other routings ((Fig~\ref{o},~\ref{p})). This is because the Epidemic routing generates unlimited number copies of each request until the request reaches the content provider. For the \textit{Epidemic} routing, the delivery ratio tends to decrease with the increase of TTL (Fig~\ref{m},~\ref{n}) and drops to 60$\%$ at TTL 500s. This is because, with a larger TTL, the \textit{Epidemic} spreads more messages which are mostly dropped by the duplicate suppression mechanism of the CIDOR and by the buffer constraint of the \textit{Epidemic} router. Fig~\ref{i} and~\ref{j} show that if a TTL value is more than 150s, it does not have any impact on the \textit{Spray-and-Wait} and the \textit{EP$_{p}$SWRouting}. The response ratio in the both \textit{Epidemic} and \textit{EP$_{p}$SWRouting} is almost 100$\%$. On the other hand, the \textit{First contact} shows a poor performance with the smaller TTL values, but performs well for the higher TTL values. In Fig~\ref{o} and~\ref{p}, we find that the average cost of the \textit{Epidemic} is quite large as compared to others.

\begin{figure*}[!t]
\centering

\subfloat[Uniform]{\includegraphics[width=1.5in]{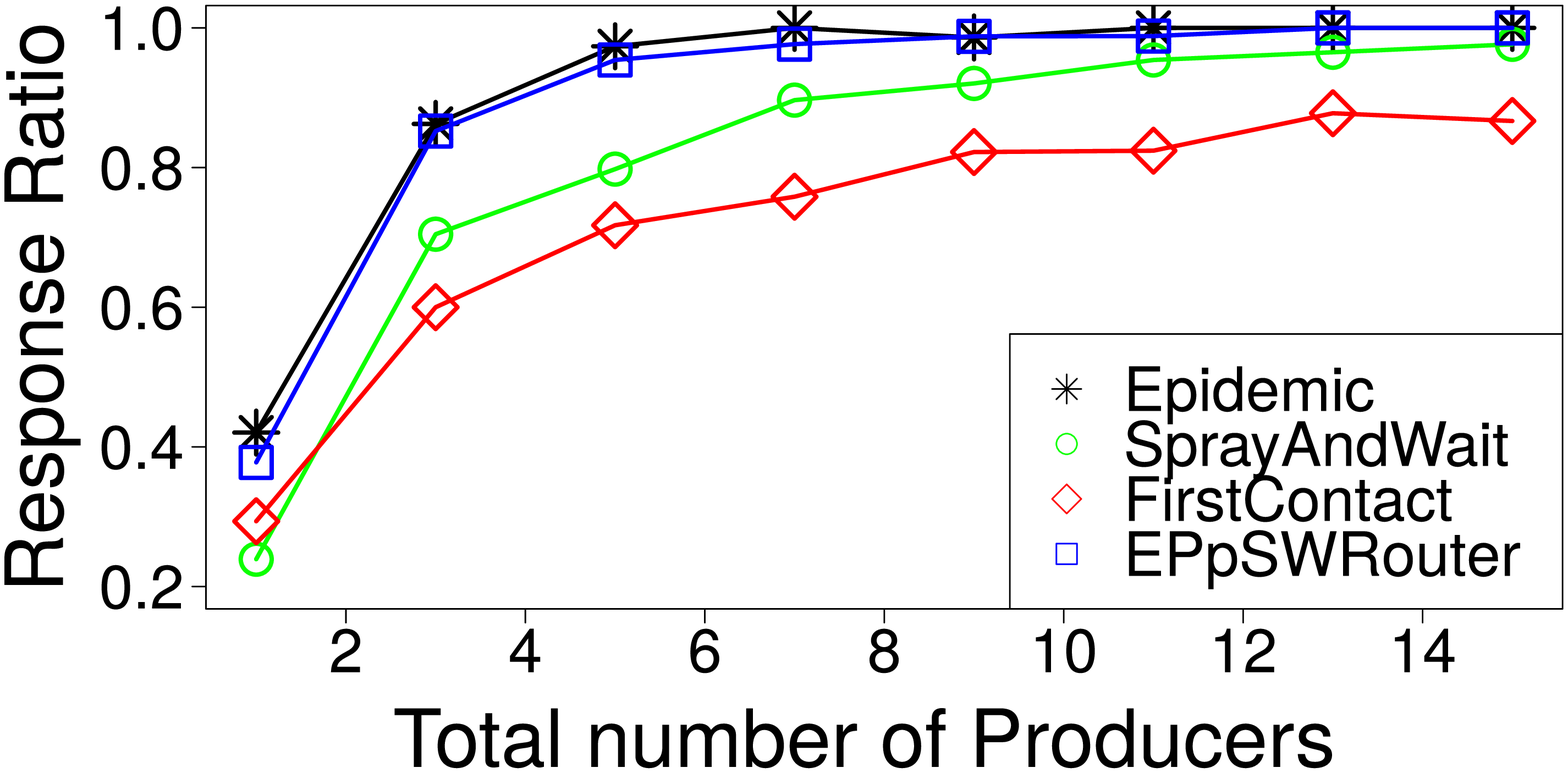}%
\label{aa}}
\hfil
\subfloat[Zipf]{\includegraphics[width=1.5in]{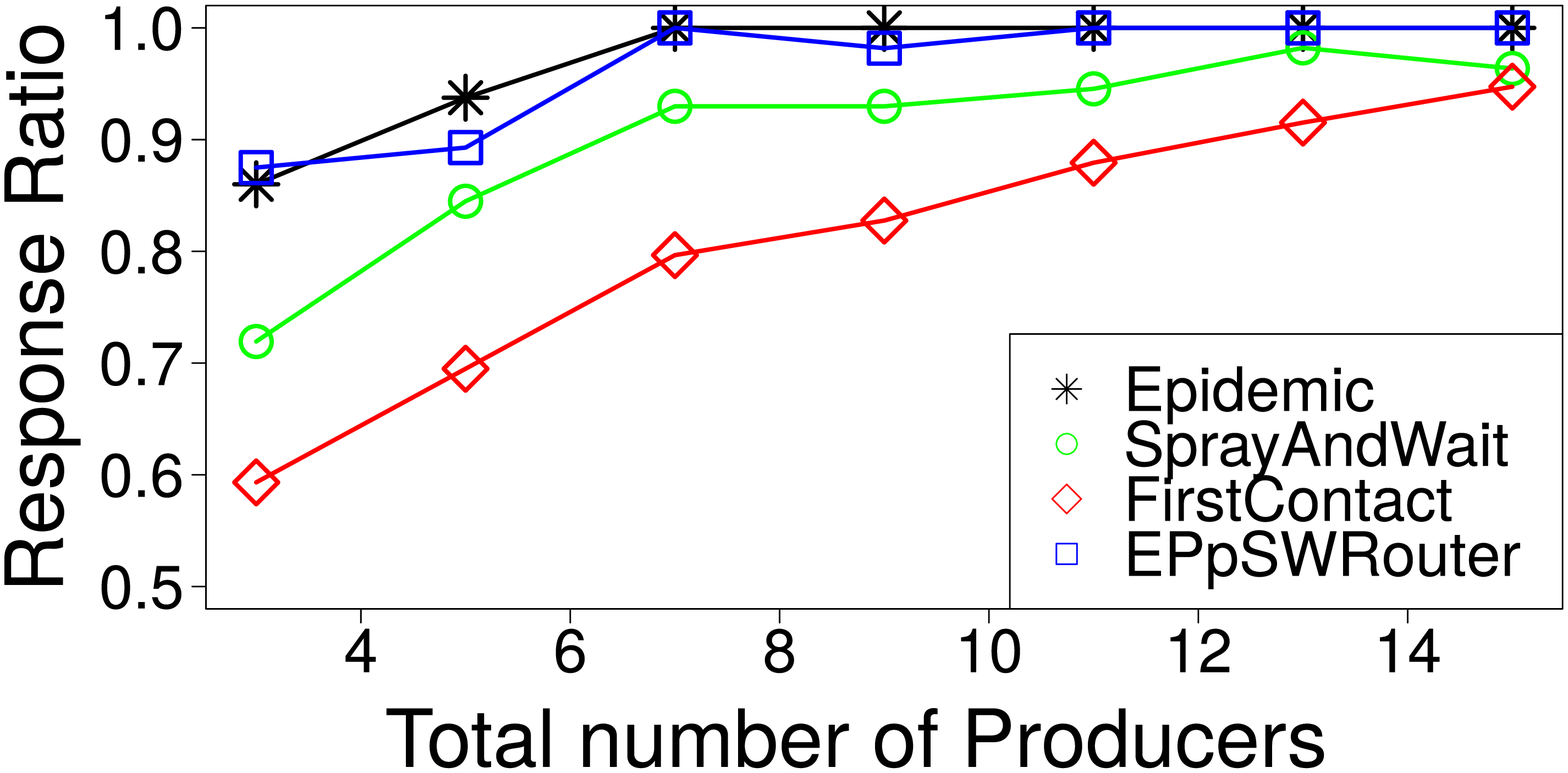}%
\label{bb}}
\hfil
\subfloat[Uniform]{\includegraphics[width=1.5in]{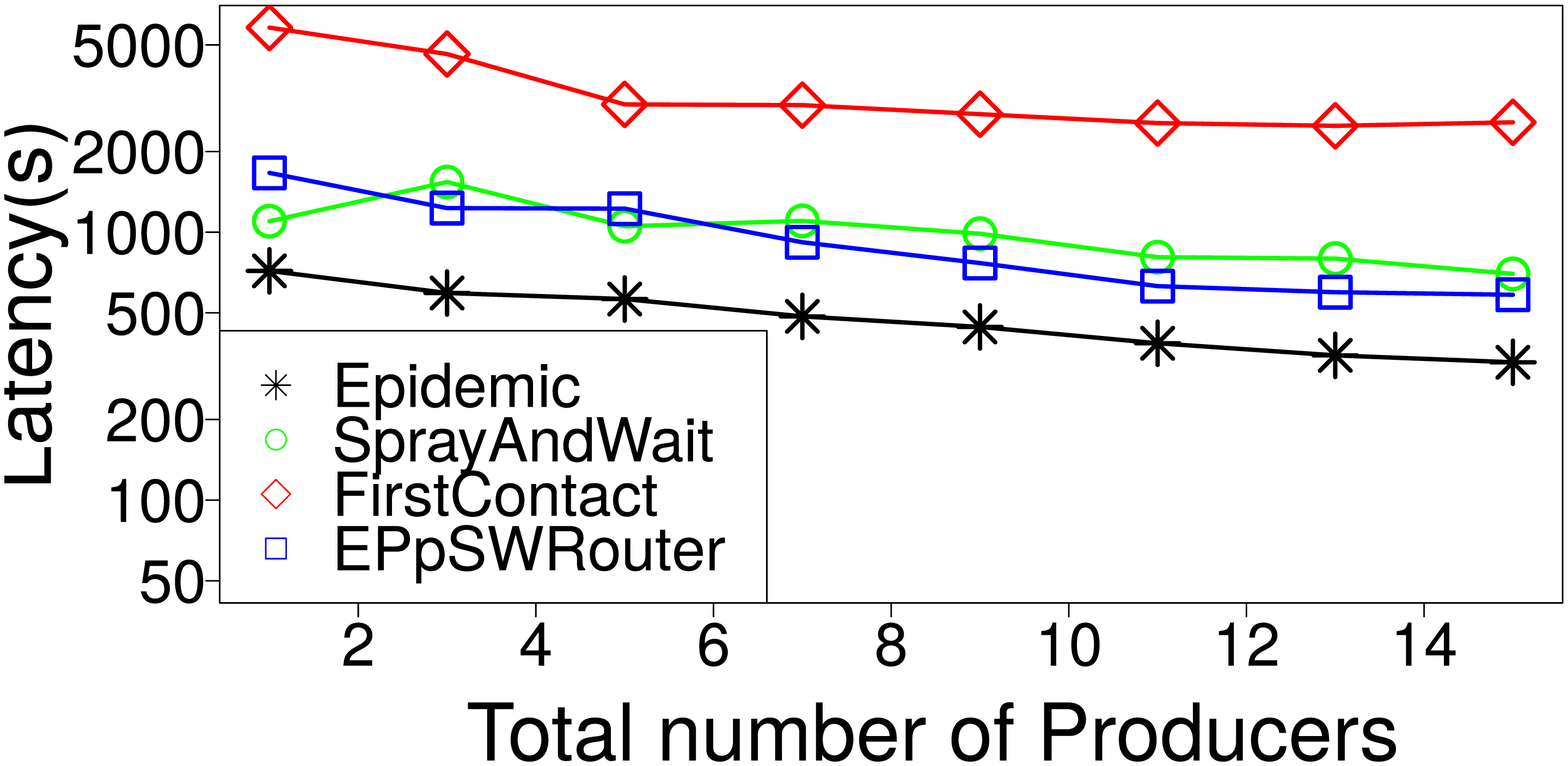}%
\label{cc}}
\hfil
\subfloat[Zipf]{\includegraphics[width=1.5in]{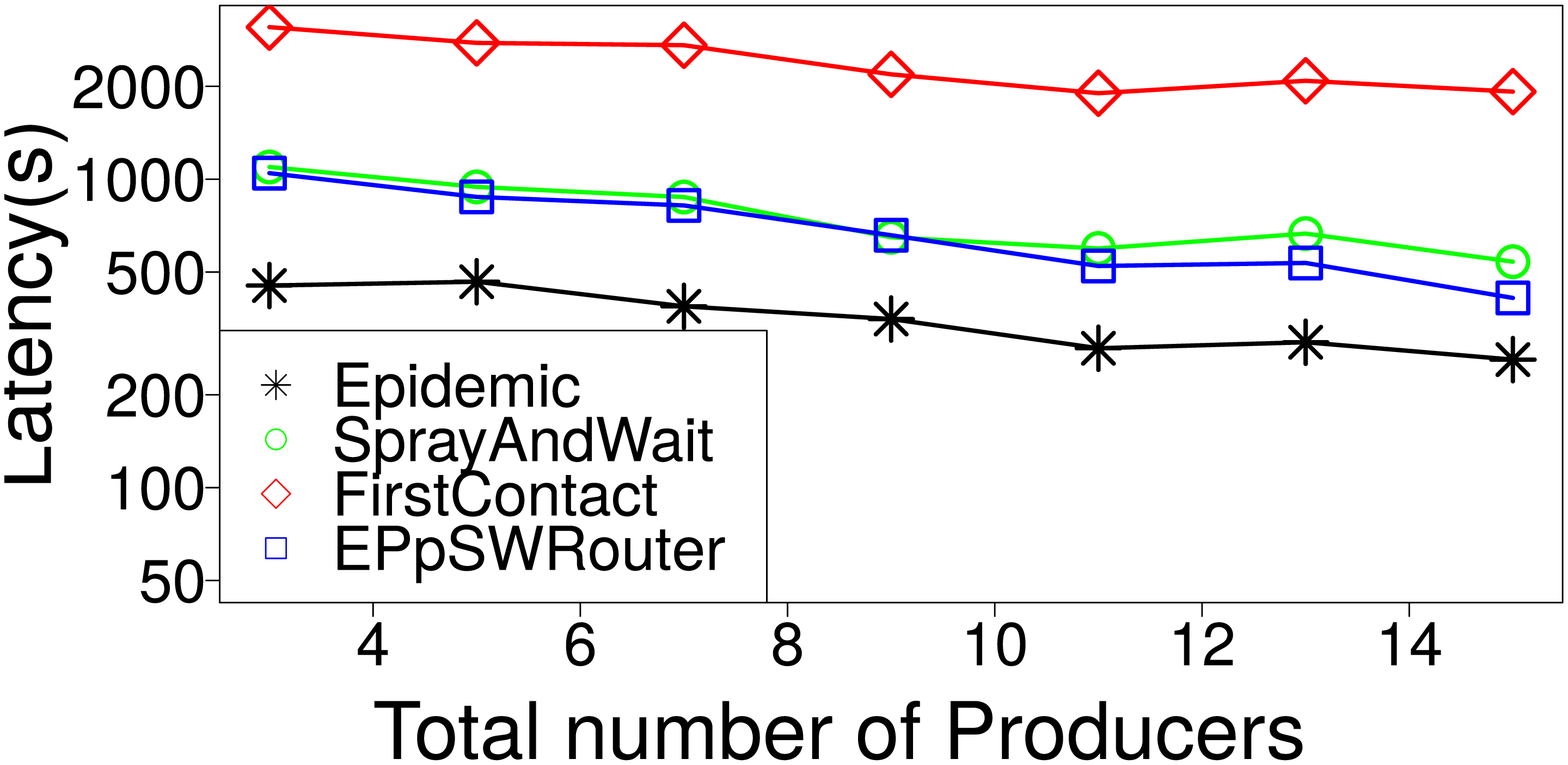}%
\label{dd}}

\vspace{.2cm}

\subfloat[Uniform]{\includegraphics[width=1.5in]{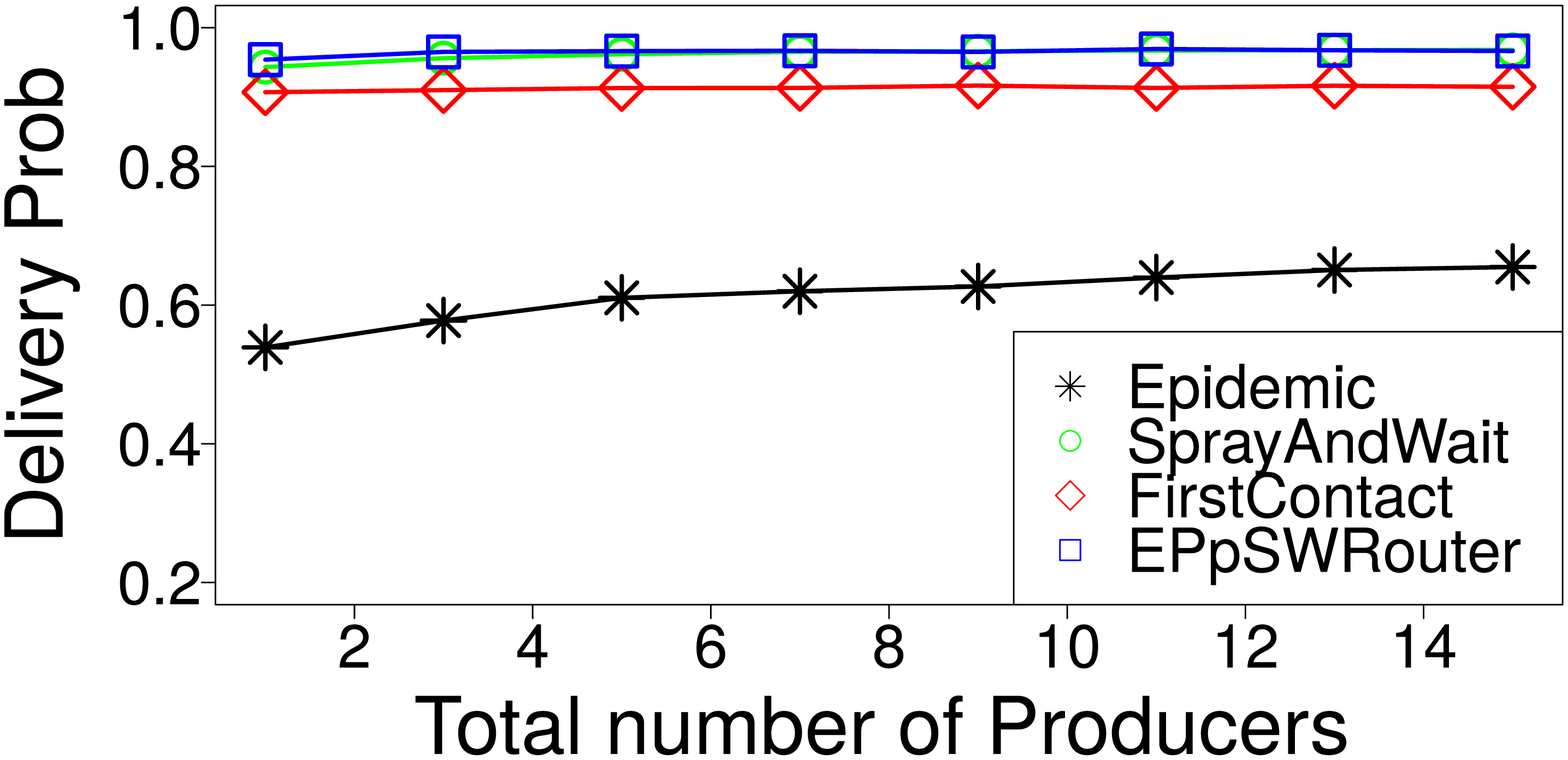}%
\label{ee}}
\hfil
\subfloat[Zipf]{\includegraphics[width=1.5in]{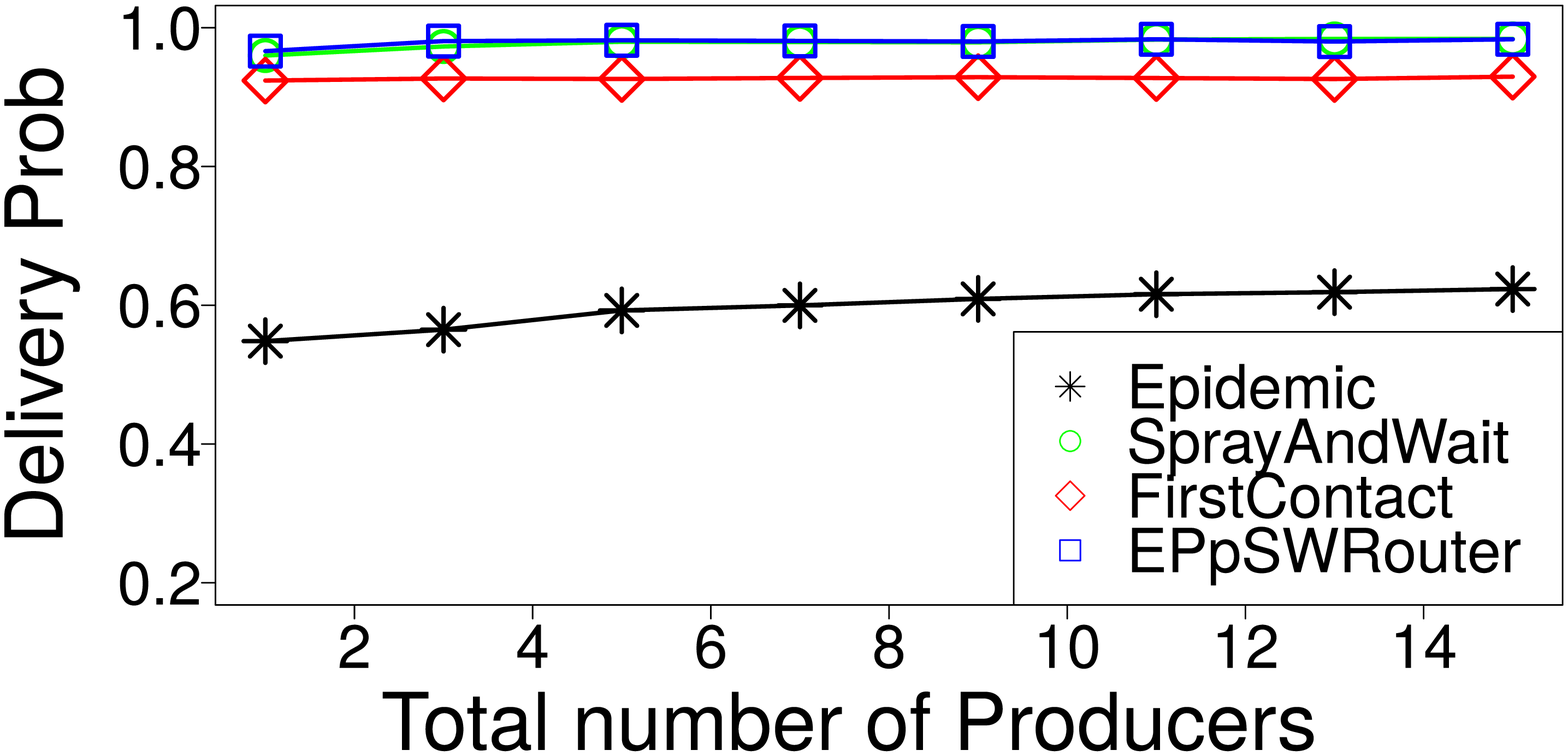}%
\label{ff}}
\hfil
\subfloat[Uniform]{\includegraphics[width=1.5in]{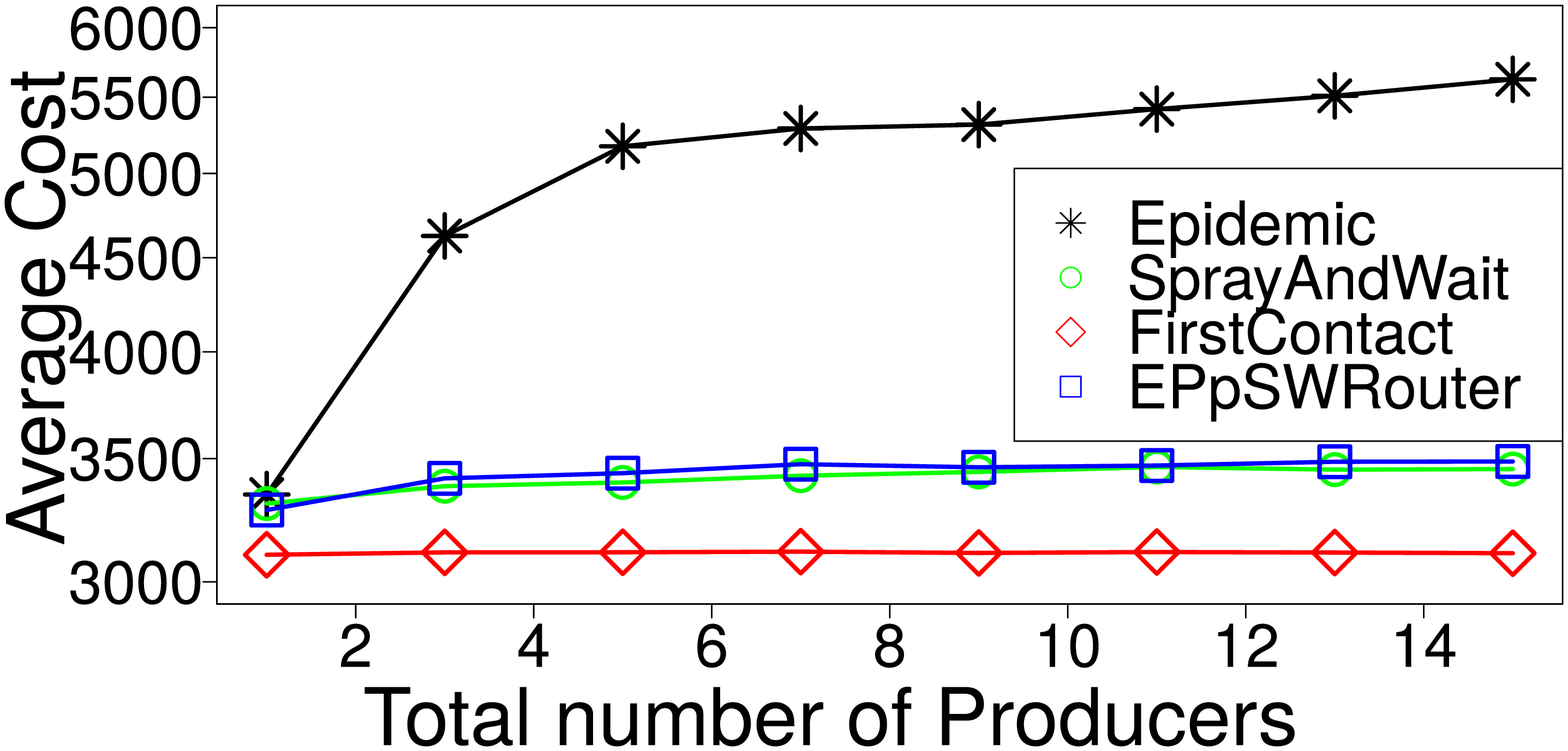}%
\label{gg}}
\hfil
\subfloat[Zipf]{\includegraphics[width=1.5in]{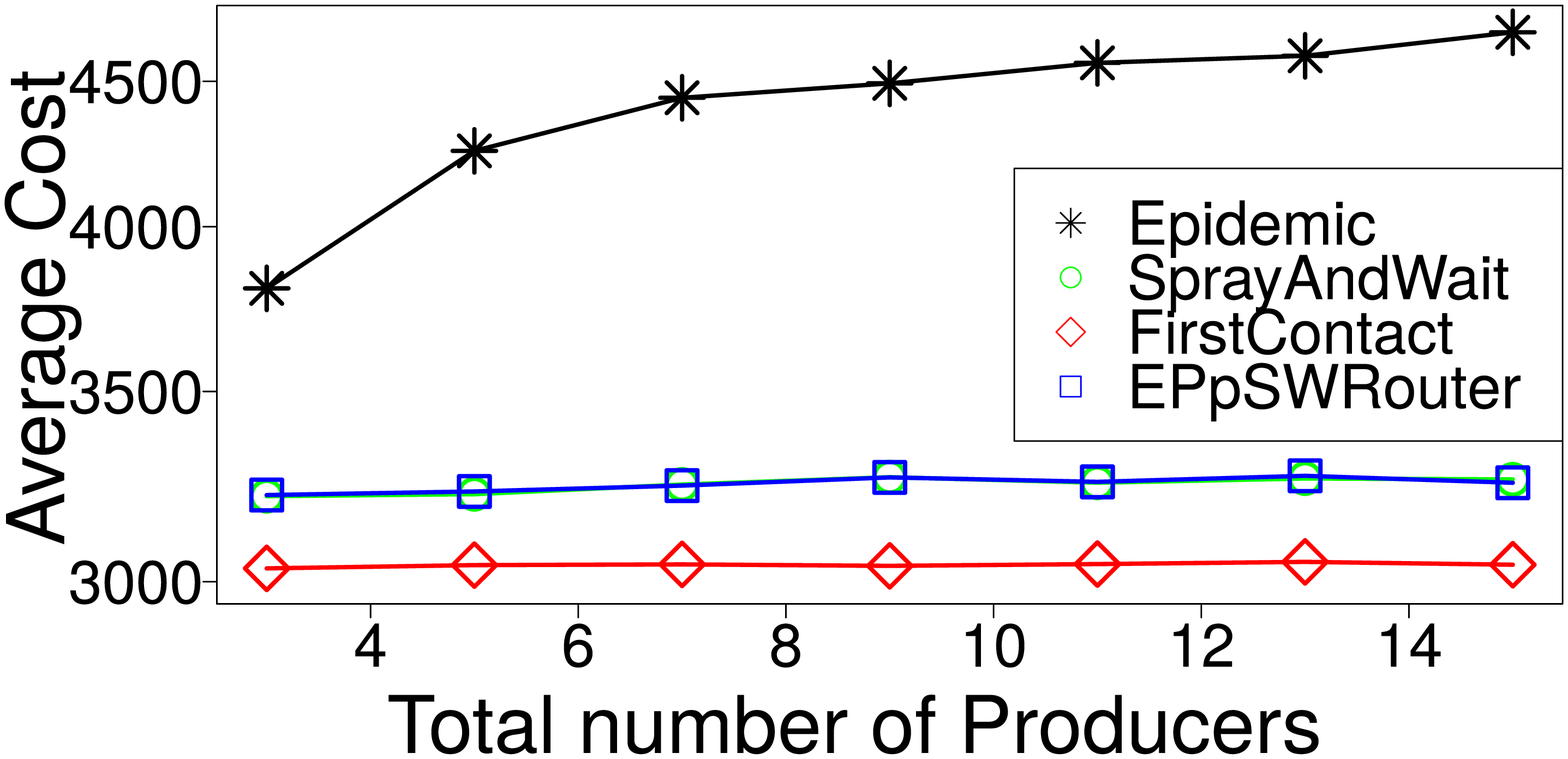}%
\label{hh}}

\caption{The behaviour of CIDOR with different DTN routing varying the number of Producers in Helsinki City Scenario.}
\label{helsinki1}
\end{figure*}

\subsection{Effect of the number of Resources}

In this experiment we fix 5 requesters, 35 intermediate nodes and vary the number of content providers. The requesters generate requests for content at a regular interval of 100s as discussed in Section~\ref{dist}. Resources are distributed to a fixed number of nodes prior to starting the simulations. Fig~\ref{aa} and~\ref{bb} show that the response ratio of all of the DTN routings gradually increase with the increase of the number of producers, as expected. Both the \textit{Epidemic} and the \textit{EP$_{p}$SWRouting} achieve the response ratio close to 100$\%$ when the number of producers are more than 8, whereas the \textit{Spray-and-Wait} routing achieves 80$\%$. The latency of retrieving content decreases with the increase of the number of producers (Fig~\ref{cc},~\ref{dd}). With 15 producers, the latency of the \textit{Epidemic}, \textit{EP$_{p}$SWRouting}, and \textit{First contact} are reduced by 67, 58, and 70 percent respectively. The delivery ratio of the \textit{Epidemic} routing ((Fig~\ref{ee},~\ref{ff})) is quite low as compared to the other routings. The \textit{Epidemic} routing gains 10$\%$ delivery ratio when the number of producers is maximum, but the average cost gradually increases ((Fig~\ref{gg},~\ref{hh})). This is because the Epidemic routing generates a unlimited number of copies of each request until the request reaches the content provider.

The interesting observation is that with the increase of the number of resources in the network the \textit{Epidemic} routing achieves 100$\%$ response ratio with a slight increase of the delivery ratio. In addition, the \textit{Spray-and-Wait} achieves a similar response ratio compared to the \textit{Epidemic} routing with a lower average cost and a comparable latency when the number of resources increases in the network.

In the RWP mobility model scenario (Fig~\ref{rwp}) we fix 10 producers and vary the number of resources in each producer. The \textit{EP$_{p}$SWRouting} and the     \textit{Epidemic} routing show a similar performance in terms of the response ratio with the increase of the number of resources (Fig~\ref{aaa},~\ref{bbb}), but the average cost of the \textit{Epidemic} routing is much higher than that of the \textit{EP$_{p}$SWRouting} (Fig~\ref{ggg},~\ref{hhh}). In terms of latency, the \textit{Epidemic} routing benefits more, as the number of resources increases. In addition, the gaps between the \textit{Epidemic} and the other routings tends to decrease (Fig~\ref{ccc},~\ref{ddd}). The delivery ratio of the \textit{Epidemic} increases by 10$\%$. The \textit{Spray-and-Wait} and the \textit{EP$_{p}$SWRouting} show a similar delivery ratio (Fig~\ref{eee},~\ref{fff}), when the number of resources are more than 30 for each producer. We observe that the delivery ratio of all of the other routings except the \textit{Epidemic} is almost constant because in this experiment we fix the buffer size to 64MB and TTL to 500s. With this settings, all the routing except \textit{Epidemic} achieves a considerably high delivery ratio (Fig~\ref{helsinki} and \ref{helsinki1}).

\subsection{Effect of Zipf Distribution}

\begin{figure*}[!t]
\centering

\subfloat[Uniform]{\includegraphics[width=1.5in]{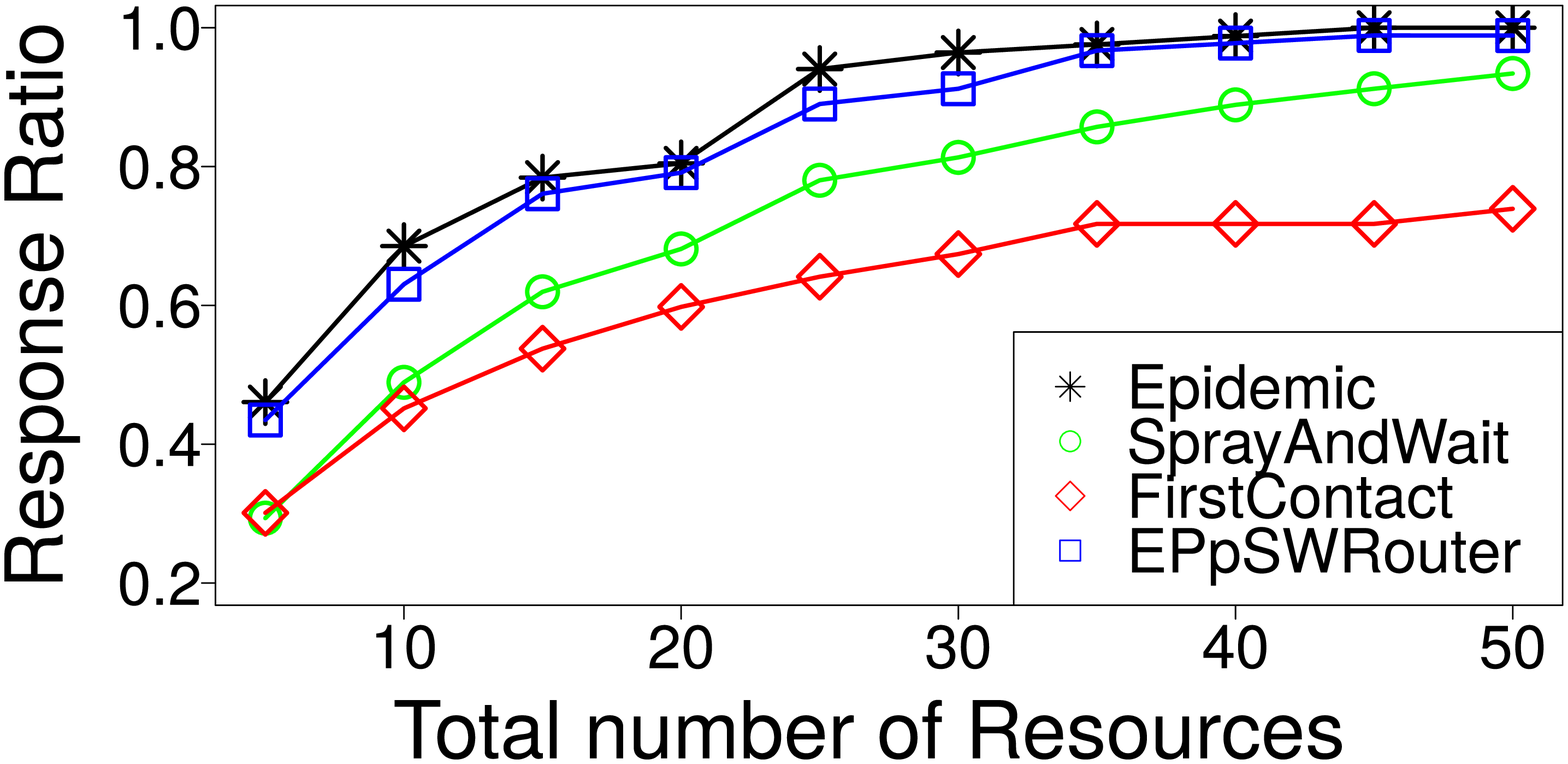}%
\label{aaa}}
\hfil
\subfloat[Zipf]{\includegraphics[width=1.5in]{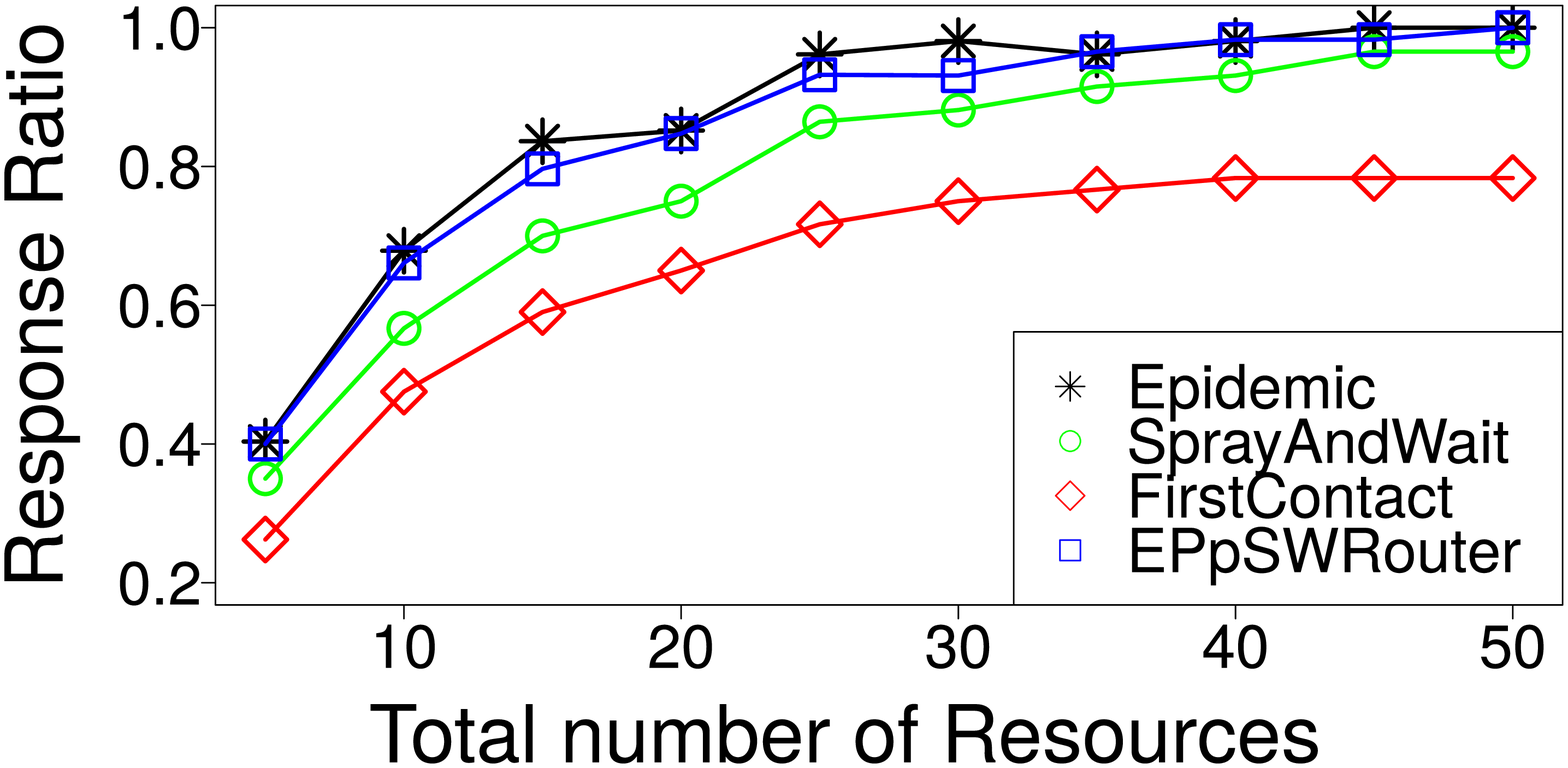}%
\label{bbb}}
\hfil
\subfloat[Uniform]{\includegraphics[width=1.5in]{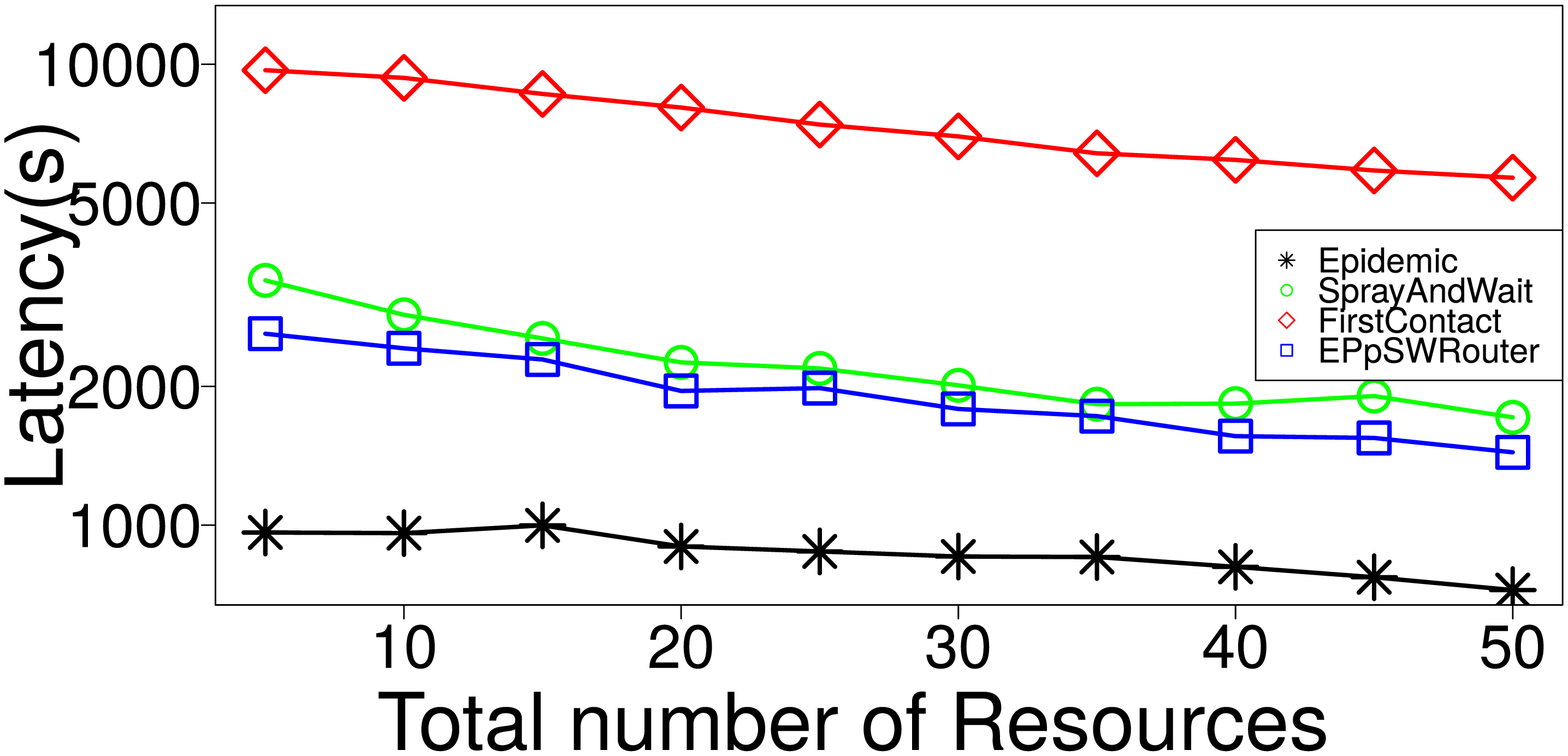}%
\label{ccc}}
\hfil
\subfloat[Zipf]{\includegraphics[width=1.5in]{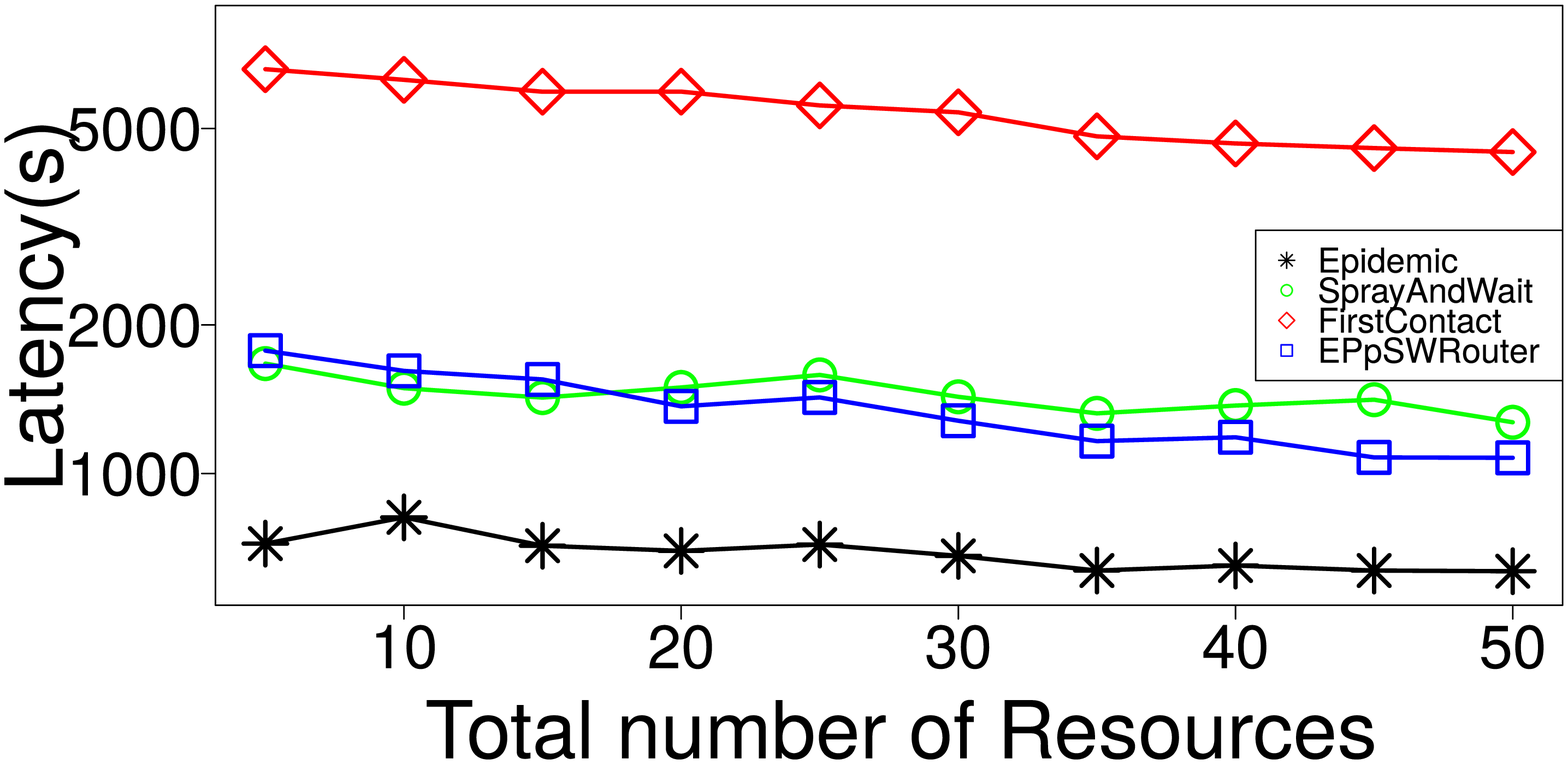}%
\label{ddd}}

\vspace{.2cm}

\subfloat[Uniform]{\includegraphics[width=1.5in]{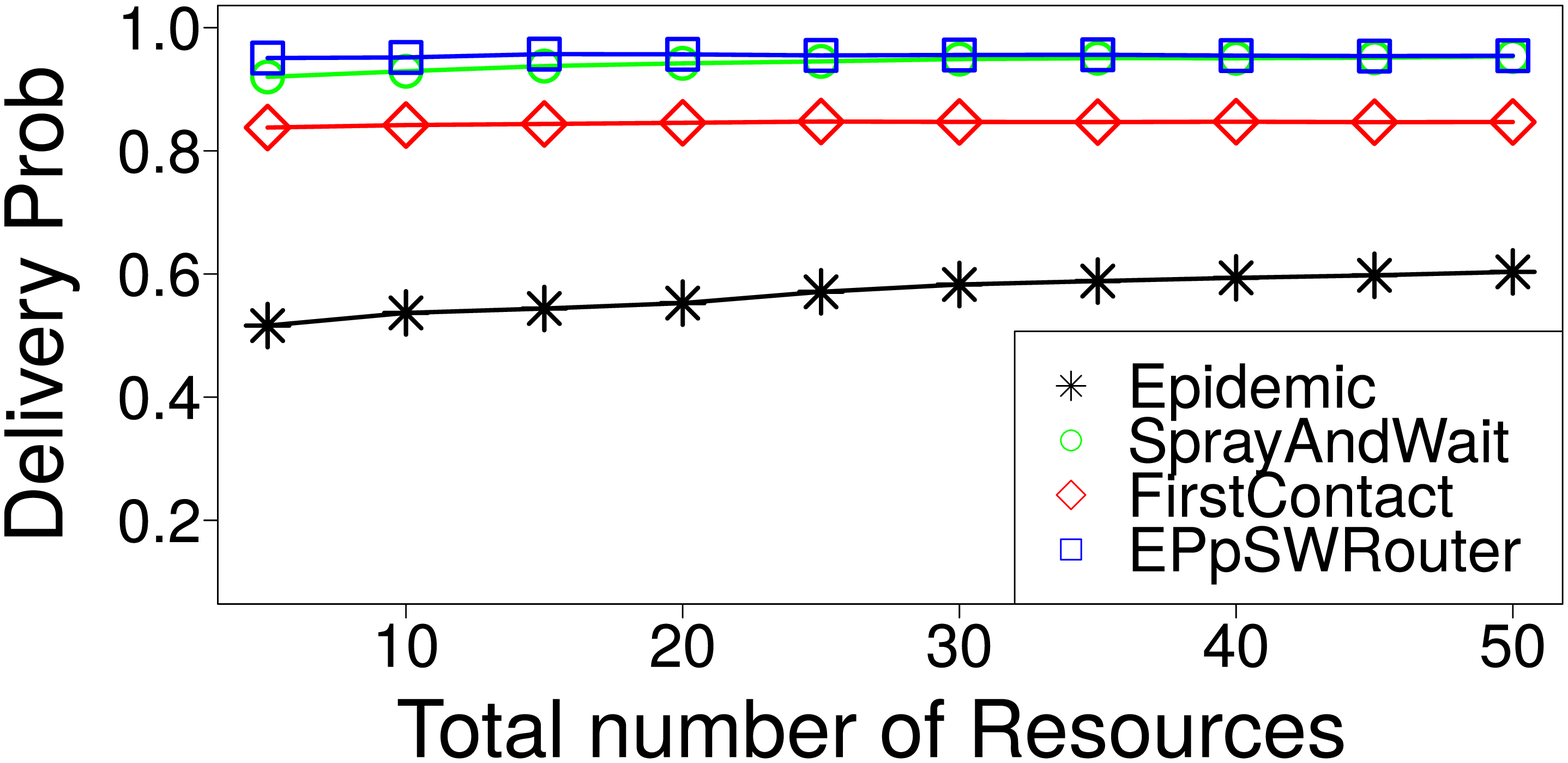}%
\label{eee}}
\hfil
\subfloat[Zipf]{\includegraphics[width=1.5in]{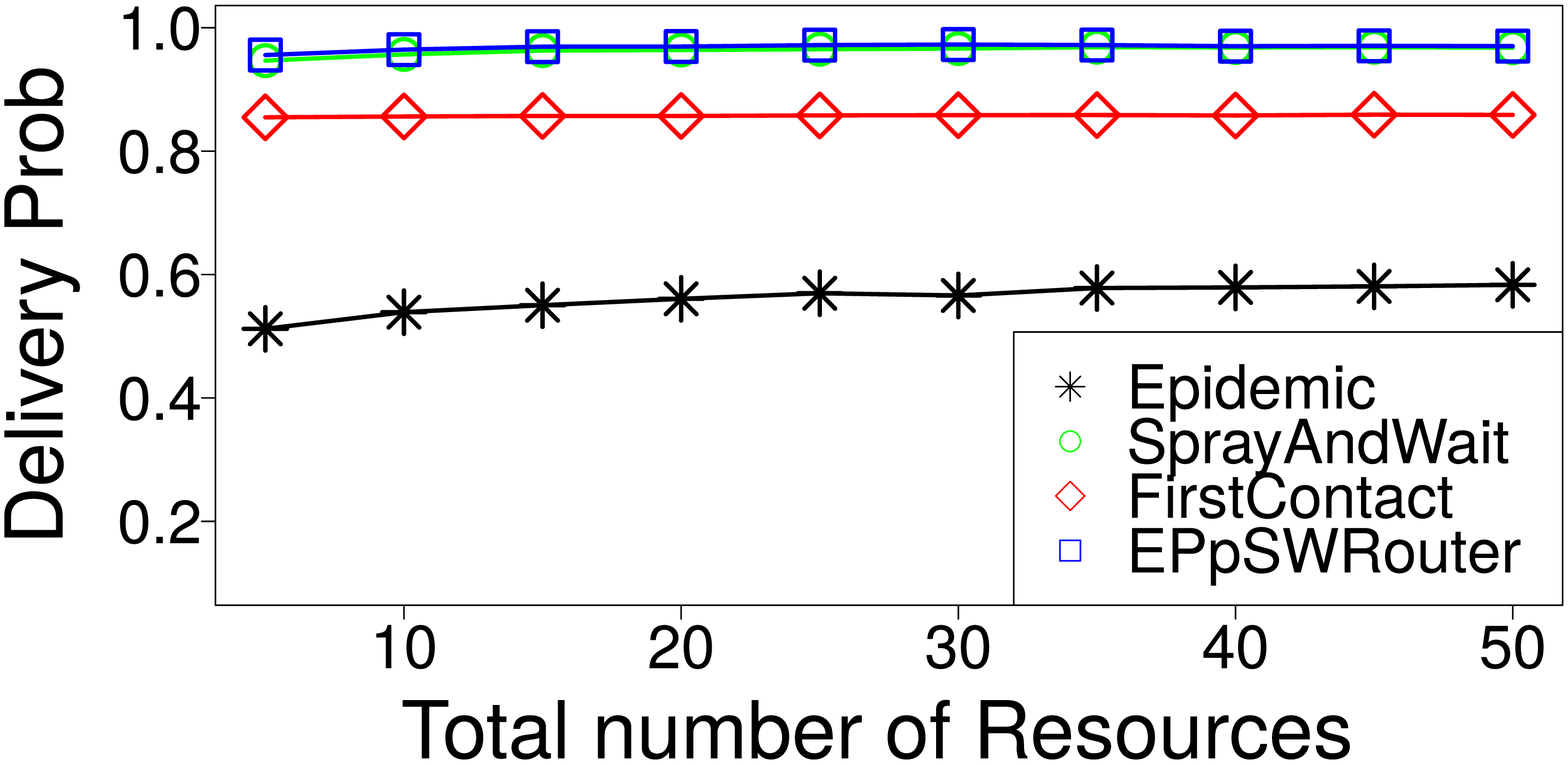}%
\label{fff}}
\hfil
\subfloat[Uniform]{\includegraphics[width=1.5in]{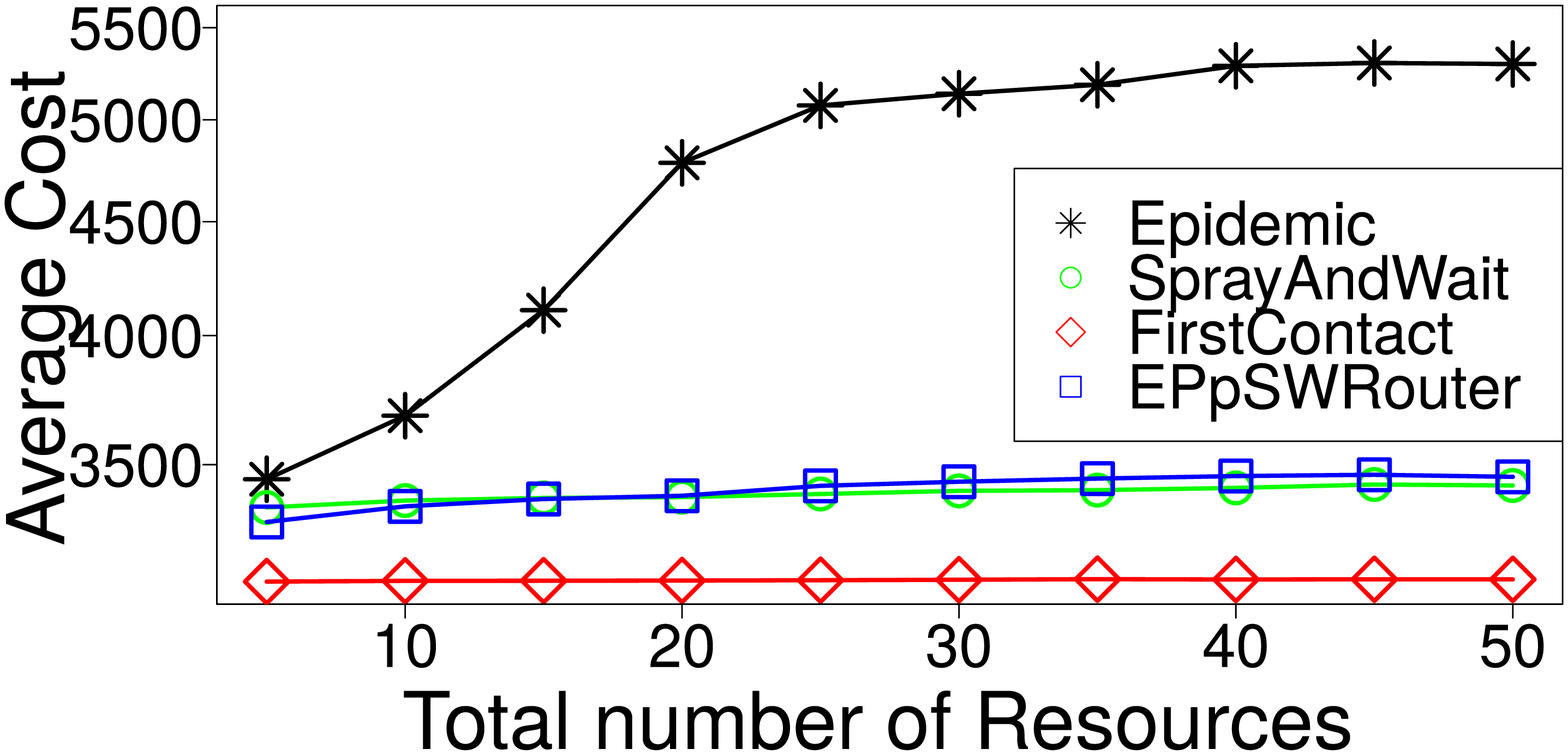}%
\label{ggg}}
\hfil
\subfloat[Zipf]{\includegraphics[width=1.5in]{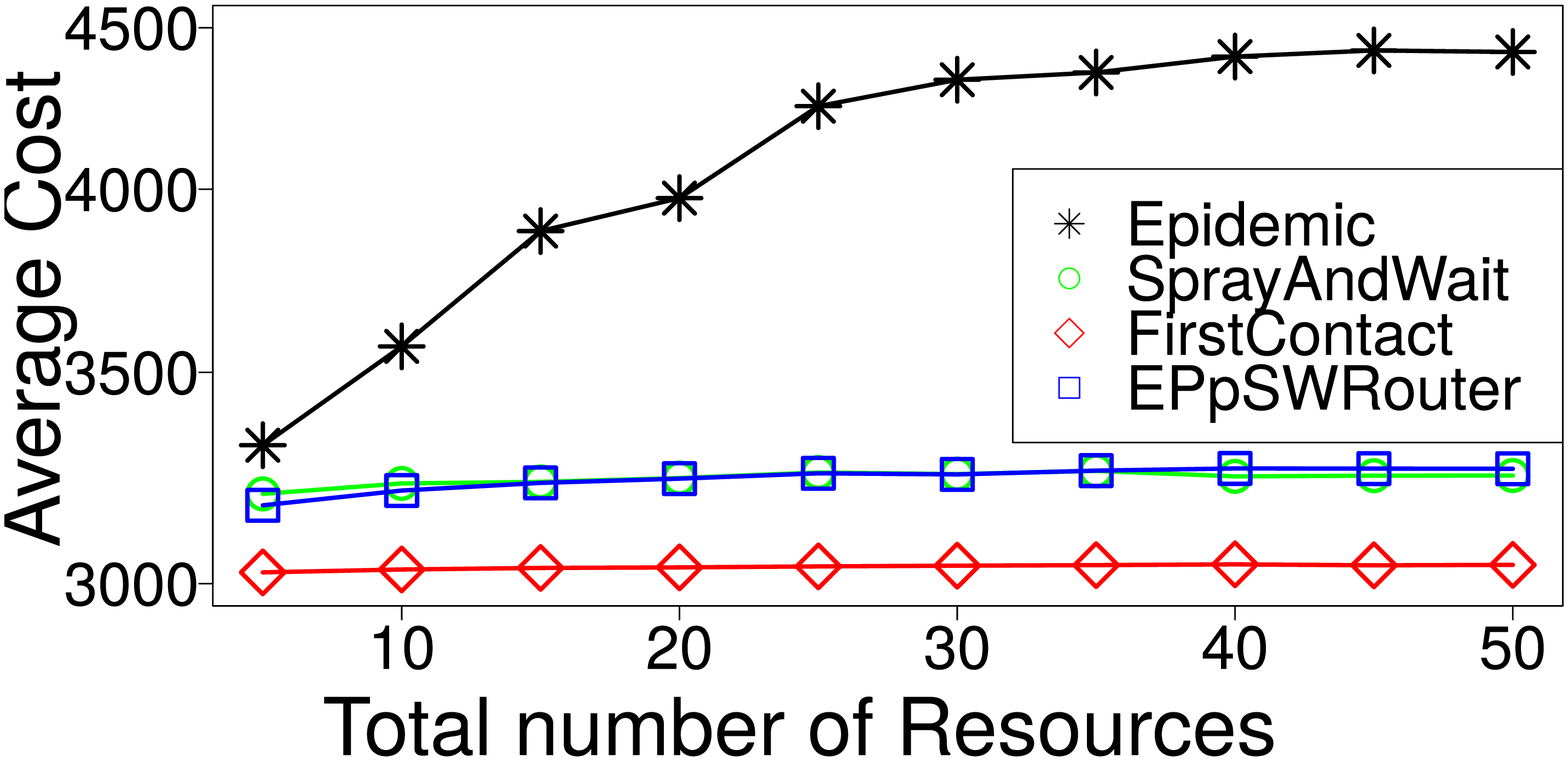}%
\label{hhh}}

\caption{The behaviour of CIDOR with different DTN routing varying the number of resources of each Producer in Random Way Point movement model.}
\label{rwp}
\end{figure*}

When the buffer size is less than 10MB, all routings in the case of the Zipf and uniform distribution show the similar performance in terms of the response ratio (Fig~\ref{b} vs. Fig~\ref{a}). With the increase of the buffer size, the \textit{First contact} and the \textit{Spray-and-Wait} gain 5$\%$ response ratio. When the buffer size is sufficiently large (above 64MB), Zipf distribution has little effect on the \textit{Epidemic} and the \textit{EP$_{p}$SWRouting}. This is because the router has sufficient space to cache both the popular and unpopular content. In terms of the average cost, the Zipf distribution has a great impact on the \textit{Epidemic} (Fig~\ref{ee} vs Fig~\ref{hh}), but a little impact on the other routings. In terms of latency, \textit{First contact} achieves good performance, since the requests for popular contents can reach the content source faster. Other routings have a little effect on latency (Fig~\ref{cc} vs Fig~\ref{dd}). In the RWP, the results show that the Zipf has a little impact on the performance of all routings in terms of response ratio and delivery probability (Fig~\ref{rwp}). However, the latency significantly drops (Fig~\ref{ccc} vs Fig~\ref{ddd}) in the case of Zipf distribution, e.g., the latency in the \textit{First contact} routing drops by 50$\%$.

\vspace{.2cm}